\documentclass[10 pt]{article}
\usepackage{setspace}
\usepackage[margin=0.8 in]{geometry}    
\usepackage[title,titletoc,toc]{appendix}
\usepackage{hyperref}     
 \usepackage{color}
 \usepackage[most]{tcolorbox}
 \tcbset{colback=yellow!10!white, colframe=red!50!black, 
        highlight math style= {enhanced, 
            colframe=red,colback=red!10!white,boxsep=0pt}
        }        
\usepackage[compat=1.1.0]{tikz-feynman}
\usepackage{graphicx}
\usepackage{amssymb}
\usepackage{amsmath}
\usepackage{amsfonts}
\usepackage{epstopdf}

\newcommand{\p}{\partial}

\newcommand{\lo}{\Lambda_0}
\newcommand{\lm}{\Lambda}
\newcommand{\hf}{{1\over 2}}
\newcommand{\be}{\begin{equation}}
\newcommand{\br}{\begin{eqnarray}}
\newcommand{\er}{\end{eqnarray}}
\newcommand{\ee}{\end{equation}}
\newcommand{\bt}{\begin{tabular}}
\newcommand{\et}{\end{tabular}}
\newcommand{\bp}{\bar p}

\newcommand{\bphi}{\bar \phi}

\newcommand{\dd}{\delta}

\newcommand{\Dp}{\frac{d^D p}{(2\pi)^D}}

\newcommand{\idp}{\int {\cal D}\phi}

\newcommand{\eps}{\epsilon}
\newcommand{\la}{\lambda}
\newcommand{\od}{\frac{1}{D}}
\newcommand{\lb}{\left\lbrace}
\newcommand{\rb}{\right\rbrace}
\newcommand{\vev}[1]{\left\langle #1 \right\rangle}
\newcommand{\N}{\mathcal{N}}
\newcommand{\ep}{\epsilon}
\newcommand{\nn}{\notag}
\newcommand{\op}[1]{\left[ #1 \right]}
\newcommand{\Op}{\mathcal{O}}

\title{Wilson Action for the $O(N)$ Model}
\author{S.~Dutta, B.~Sathiapalan\\
Institute of
  Mathematical Sciences\\CIT Campus, Tharamani\\ 
  Chennai 600113, India\\and\\Homi Bhabha National Institute\\Training
  School Complex, Anushakti Nagar\\Mumbai 400085, India\\
and\\\\H.~Sonoda
\\Physics Department, Kobe
  University\\Kobe 657-8501, Japan}

\begin{document}

\hspace{12cm} IMSc/2020/02/02\\

\hspace{12cm} KOBE-TH-20-02
{\let\newpage\relax\maketitle}

\maketitle
\begin{abstract}
  In this paper the fixed-point Wilson action for the critical $O(N)$
  model in $D=4-\eps$ dimensions is written down in the $\eps$
  expansion to order $\eps^2$. It is obtained by solving the
  fixed-point Polchinski Exact Renormalization Group equation (with
  anomalous dimension) in powers of $\eps$. This is an example of a
  theory that has scale and conformal invariance despite having a
  finite UV cutoff. The energy-momentum tensor for this theory is also
  constructed (at zero momentum) to order $\eps^2$. This is done by
  solving the Ward-Takahashi identity for the fixed point action. It
  is verified that the trace of the energy-momentum tensor is
  proportional to the violation of scale invariance as given by the
  exact RG, i.e., the $\beta$ function.  The vanishing of the trace at
  the fixed point ensures conformal invariance. Some examples of 
  calculations of correlation functions are also given.
\end{abstract}
\newpage
\tableofcontents
\newpage

\section{Introduction}

Conformal field theories (CFT) are interesting for a variety of
reasons. One of the most important reason is that a theory critical at
a continuous phase transition is expected to acquire conformal
invariance which imposes strong constraints on the correlation
functions\cite{Polyakov1970}. This has motivated the idea of
bootstrap\cite{Polyakov1974}. Particularly in two dimensions these
ideas have been very fruitful \cite{Polyakov1984}. Reviews of later
developments and references are given in
\cite{DiFrancesco1997,Rychkov2016}.

The advent of the AdS/CFT correpondence
\cite{Maldacena,Polyakov,Witten1,Witten2} or ``holography'' between a
boundary CFT and a bulk gravity theory opened up another approach to
solving CFT's. \footnote{It also opens up the amazing possibility of
  rewriting quantum gravity as a quantum field theory in flat space.}
There is a large amount of literature on this.  See, for example,
\cite{Penedones2016} for a review.

In the AdS/CFT correspondence the radial direction can be interpreted
as the scale of the boundary field theory. Thus, a radial evolution
can be thought of as an RG evolution and has been dubbed ``holgraphic
RG''
\cite{Akhmedov,Akhmedov1,Akhmedov2,Alvarez,Kraus,Warner,Verlinde,Boer,Faulkner,Klebanov:1999tb,Heemskerk,Morris,Bzowski:2015pba,deHaro:2000vlm}.
The precise connection between the boundary RG and holographic RG is,
however, still an open question.

Recently a connection has been proposed between the Exact
Renormalization Group (ERG) equation
\cite{Wilson,Wegner,Wilson2,Polchinski} and the Holographic
Renormalization Group (Holographic RG) equation. It was shown in
\cite{Sathiapalan:2017frk} that the RG evolution operator for a Wilson
action of a D-dimensional field theory obeying the Polchinski ERG
equation can be formulated as a $D+1$-dimensional functional
integral. The extra dimension, corresponding to the moving scale $\lm$
of the ERG, makes it a ``holographic'' formulation. Furthermore, a
change of field variables or field redefinition maps the $D+1$
dimensional action for the functional integral to the action of a free
massive scalar field in $AdS_{D+1}$.  It was then shown that the
calculation of the two point function reduces to the familiar
calculation using the AdS/CFT correspondence.

This proposal is quite general, and detailed calculations were done
for the Gaussian theory \cite{Sathiapalan:2017frk}.  The scalar field
theory action has a free parameter, i.e., the mass of the scalar
field, which is related to the anomalous dimension of the boundary
operator in the AdS/CFT context .  This parameter appears to come out
of nowhere.  To understand the origin of the anomalous dimension
parameter, an ERG equation with anomalous dimension was analysed in
\cite{Sathiapalan:2019zex}.  The same change of variables mapped this
to a scalar field theory in the AdS space-time, and this time it was
easy to see that the mass parameter is naturally related to the
anomalous dimension parameter in the ERG. Normally, interactions are
required for a field to have anomalous dimension. Since the exact RG
for interacting theories is difficult, a Gaussian theory with an
anomalous dimension introduced by hand was studied in
\cite{Sathiapalan:2019zex}.

In order to improve our understanding of the connection between ERG
and the AdS/CFT correspondence, it is necessary to have an interacting
example --- one needs a non-trivial boundary CFT and a fixed-point
Wilson action for this CFT \footnote{Note that the ``Wilson action''
  always has a finite UV cutoff --- this is a point of departure from
  the usual CFT actions written in the continuum.}.  Then the RG
evolution of small perturbations to this theory can be studied by ERG.
Using the ideas of \cite{Sathiapalan:2017frk,Sathiapalan:2019zex} this
can be mapped to a scalar field theory in $D+1$-dimensional AdS
space. This would make a contact with more detailed AdS/CFT
calculations of higher point correlators.  A well studied field theory
is the $\la \phi^4$ scalar field theory in $4-\eps$ dimensions that
has the famous Wilson-Fisher fixed point. When there are $N$ scalar
fields, this is often referred to as the $O(N)$ model. In this paper,
as a first step, we construct a fixed-point Wilson action for this
theory to order $\eps^2$. It is at this order that the anomalous
dimension first shows up.  The action is obtained by solving the
fixed-point ERG equation perturbatively. The fixed-point equation
imposes the constraint of scale invariance. 

In fact the theory is also
conformally invariant. This follows from the properties of the energy
momentum tensor --- if it is traceless the theory is conformally
invariant. Indeed the tracelessness of the energy-momentum tensor
defines what we mean by a CFT \cite{Callan:1970ze, Coleman2,Brown,
  Polchinski2}. It is thus important to study the energy-momentum tensor and we construct it in this paper.
 
The energy-momentum tensor is also important in the context of
AdS/CFT: one of the really interesting aspects of the AdS/CFT
correspondence is that the $D+1$-dimensional bulk theory has dynamical
gravity. In addition to the scalar field, there is the gravitational
field that couples to the energy momentum tensor of the boundary CFT. Thus to extend the ideas
of \cite{Sathiapalan:2017frk,Sathiapalan:2019zex} to understand bulk gravity in AdS/CFT correspondence, from ERG
one has to construct the energy momentum operator.

 The energy-momentum tensor for $\phi^4$ field theory
has been worked out in the dimensional regularization scheme
\cite{Brown}.  The construction of the energy-momentum tensor from the
ERG point of view has been studied in general in
\cite{Sonoda-emt,Rosten2}. The main idea is to solve the Ward Identity associated with
coordinate transformations. This can be done in perturbation theory. We construct the leading terms that corresponds to the zero momentum 
energy momentum tensor. One can also check that the trace
of the energy momentum tensor is proportional to the number operator. We apply this prescription here and
construct the zero momentum energy momentum tensor to $O(\la^2)$.


This paper is organized as follows: In Section 2 we give a review of
ERG and the fixed-point equation. We also give some background
material on the energy-momentum tensor.  In Section 3 we construct the
solution to the fixed-point equation and obtain the fixed-point
action. In Section 4 we give a different approach to obtaining the fixed point equation and also calculate some correlation functions. In Section 5 the construction of the energy-momentum tensor is
given. We conclude the paper in Section 6.

\section{Background}
\label{secA}
\subsection{Exact Renormalization Group and Fixed Point equation}
\label{secA1}

We review the necessary background in this section. It depends mostly
on \cite{Igarashi, Sonoda-equiv}.

\subsubsection{ Exact Renormalization group}

\vspace{0.1 in} Renormalization means essentially going from one scale
$\Lambda_0$ to a lower scale $\Lambda$, where the initial scale
$\Lambda_0$ is typically called a bare scale. One will want to see how
the physics changes with scale. What do we mean by physics at
$\Lambda_0$? It means our theory will not be sensitive to momentum
$p >\Lambda_0$. The partition function of the full theory is given by
\begin{align*}
Z= \int\mathcal{D}\phi \, e^{-S[\phi]}
\end{align*}
where
\begin{align*}
S= \int_{p}\frac{1}{2} p^2 \phi^2 +S_{I}[\phi]
\end{align*}
To make it a partition function at scale $\Lambda_0$ we will try to
suppress the kinetic energy term for
$\infty < p < \Lambda_0 $. To execute this we will put a smooth cutoff
in the kinetic energy term to obtain the bare action
\begin{align}
S_B [\phi] \equiv \frac{1}{2}\int_p
  \phi\frac{p^2}{K(p^2/\Lambda_0^2)} \phi  + S_{I,B} [\phi]
\end{align}
and the bare partition function
\begin{equation}
Z_B \equiv \int \mathcal{D}\phi\, e^{- S_B [\phi]}
\end{equation}

We will choose the cutoff function will follow the condition \hspace{0.05 in}$K(0)=1$ and $K(\infty)=0$. In general cutoff functions satisfy stronger properties , but that will not affect the fixed point values of the couplings \cite{Igarashi-gamma}.

\vspace{0.1 in}

\vspace{0.02 in} Now we want to go to a lower scale $\Lambda$. For
that, observe the following identity
\begin{align*}
&\int \mathcal{D}\phi \exp \left[ -\frac{1}{2} \int_p
  \phi(-p)\frac{1}{A(p)+B(p)}\phi(p)-S_{I,B}[\phi]\right]\\ 
=& \int \mathcal{D}\phi_1 \mathcal{D} \phi_2 \exp
   \left[-\frac{1}{2}\int_p
   \frac{1}{A(p)}\phi_1(-p)\phi_1(p)-\frac{1}{2}\int_p
   \frac{1}{B(p)}\phi_2(-p)\phi_2(p)- S_{I,B}[\phi_1+\phi_2]\right] 
\end{align*}
Using this we can write
\begin{align}
\nonumber Z_B= &\int \mathcal{D}\phi_l \mathcal{D} \phi_h \exp \bigg
                 \lbrace-\frac{1}{2}\int_p
                 \frac{p^2}{K(p^2/\Lambda^2)}\phi_l(-p)\phi_l(p)\\
  \nonumber 
-&\frac{1}{2}\int_p
   \frac{p^2}{K(p^2/\Lambda_0^2)-K(p^2/\Lambda^2)}\phi_h(-p)\phi_h(p)-
   S_{I,B}[\phi_l+\phi_h]\bigg \rbrace 
\end{align}
We can effectively call $\phi_l(\phi_h)$ as low(high) energy field as
it is propagated by low(high) momentum propagator $\Delta_l(\Delta_h)$ defined below 
\begin{align}\label{prop}
\Delta_l= \frac{K(p^2/\Lambda^2)}{p^2},\quad
  \Delta_h=\frac{K(p^2/\Lambda^2)-K(p^2/\Lambda_0^2)}{p^2} 
\end{align}
So we can write
\begin{align*}
Z_B= &\int \mathcal{D}\phi_l \exp \left[ -\frac{1}{2}\int_p \phi_l
       \Delta_l^{-1} \phi_l\right] \int \mathcal{D}\phi_h \exp
       \left[-\frac{1}{2} \int_p \phi_h \Delta_h^{-1} \phi_h- S_{I,B}[\phi_l+\phi_h]\right]\\
=& \int \mathcal{D} \phi_l \exp \left[ -\frac{1}{2}\int_p \phi_l
   \Delta_l^{-1} \phi_l\right] \exp \lbrace -S_{I,\Lambda}
   [\phi_l]\rbrace
\end{align*}
where 
\begin{align}\label{SILambda}
\exp \lbrace -S_{I,\Lambda}[\phi_l] \rbrace \equiv \int
  \mathcal{D}\phi_h \exp\bigg \lbrace
  -\frac{1}{2}\int_p \phi_h\Delta_h^{-1}\phi_h-S_{I,B}[\phi_l+\phi_h]\bigg
  \rbrace 
\end{align}
$S_{I, \Lambda}$ is the interaction part of an effective low energy
field theory with a UV cutoff $\Lambda$.
\newline
Let 
\begin{equation}\label{SLambda}
S_\Lambda[\phi] \equiv \frac{1}{2} \int_p \phi_l \Delta_l^{-1}
  \phi_l+ S_{I,\Lambda} [\phi_l]
\end{equation}
be the whole action so that
\begin{equation}
Z_B = \int \mathcal{D} \phi_l\, e^{- S_\Lambda [\phi_l]}
\end{equation}
Using (\ref{SILambda}), we obtain
\begin{align}
e^{- S_\Lambda [\phi]} 
&= \int \mathcal{D}\varphi\, \exp \left[ - S_B [\varphi] + \frac{1}{2}
  \int_p \frac{p^2}{K(p/\Lambda_0)} \varphi (p) \varphi (-p)
 - \frac{1}{2} \int_p \frac{p^2}{K(p/\Lambda)} \phi (p)
  \phi (-p) \right.\notag\\
&\left.\quad - \frac{1}{2} \int_p \frac{p^2}{K(p/\Lambda_0) -
  K(p/\Lambda)} \left( \varphi (p) - \phi (p)\right)\left(\varphi
  (-p)-\phi (-p)\right) \right]\label{SB-SLambda}
\end{align}
where we have written $\phi_l$ as $\phi$ and $\phi_h$ as $\varphi - \phi$.
This will be useful later.

\vspace{0.1 in}

%

It is to be noted that one can go back to the bare partition function
anytime . For this reason this scheme is called \textbf{``exact''},
i.e. we lose no physical information by varying the scale.  It is
easy to see this explicitly.  Using (\ref{SB-SLambda}), we can
calculate the generating functional of $S_B$ using $S_\Lambda$ as
\begin{align}
&\int \mathcal{D} \phi\, \exp \left( - S_B [\phi] - \int_p J(-p) \phi
  (p) \right)\notag\\
&=  \exp \left[ \frac{1}{2} \int_p J(p) J(-p) \frac{1}{p^2} \left\lbrace
  K(p/\Lambda_0)\left(1 - K(p/\Lambda_0)\right) -
  \left(\frac{K(p/\Lambda_0)}{K(p/\Lambda)}\right)^2
  K(p/\Lambda)\left(1 - K(p/\Lambda)\right) \right\rbrace
  \right]\notag\\
&\qquad \times \int \mathcal{D} \phi\, \exp \left( - S_\Lambda [\phi] - \int_p
  J(-p) \frac{K(p/\Lambda_0)}{K(p/\Lambda)}\, \phi (p)\right)
\label{ZBJ-SL}
\end{align}
We observe that the correlation functions of $S_B$ are the same as
those of $S_\Lambda$ up to the trivial (short-distance) contribution
to the two-point function and up to the momentum-dependent rescaling
of the field by $\frac{K (p/\Lambda_0)}{K (p/\Lambda)}$
\cite{Sonoda-equiv}.   If we ignore the small corrections to the
two-point functions, we can write
\begin{equation}
\prod_{i=1}^n \frac{1}{K(p_i/\Lambda)}\, \vev{\phi (p_1) \cdots \phi
  (p_n)}_{S_\Lambda}
= \prod_{i=1}^n \frac{1}{K(p_i/\Lambda')}\, \vev{\phi (p_1) \cdots \phi
  (p_n)}_{S_{\Lambda'}}\label{SL-correlations}
\end{equation}

\subsubsection{Polchinski's ERG equation}

We have given an integral formula (\ref{SILambda}) for $S_{I,\Lambda}$ and
(\ref{SB-SLambda}) for $S_\Lambda$.  It is easy to derive differential equations
from these.  From (\ref{SILambda}), we obtain Polchinski's ERG
equation
\begin{equation}\label{polpsi}
- \Lambda \frac{\partial S_{I,\Lambda} [\phi]}{\partial \Lambda}
= \int_p (-) \frac{d K(p/\Lambda)}{dp^2} \left(-
\frac{\delta S_{I,\Lambda}[\phi]}{\delta \phi (p)}\frac{\delta
  S_{I,\Lambda} [\phi]}{\delta \phi (-p)} + \frac{\delta^2
  S_{I,\Lambda} [\phi]}{\delta \phi (p) \delta \phi (-p)} \right)
\end{equation}
for $S_{I,\Lambda}$.  From (\ref{SB-SLambda}) we obtain
\begin{equation}
- \Lambda \frac{\partial S_\Lambda [\phi]}{\partial \Lambda}
= \int_p \left[ - 2 p^2 \frac{d \ln K(p/\Lambda)}{dp^2} \, \phi (p)
  \frac{\delta S_\Lambda}{\delta \phi (p)} + \frac{d K(p/\Lambda)}{dp^2} \left(
    - \frac{\delta S_\Lambda}{\delta \phi (p)} \frac{\delta
      S_\Lambda}{\delta \phi (-p)} + \frac{\delta^2 S_\Lambda}{\delta
      \phi (p) \delta \phi (-p)} \right)\right]\label{ERG-SL}
\end{equation}
for the entire Wilson action.

\subsubsection{The limit $\Lambda \to 0+$}

In the limit $\Lambda \to 0+$ we expect $S_\Lambda [\phi]$ approaches something related
to the partition function.  If we substitute
\begin{equation}
\lim_{\Lambda \to 0+} K(p/\Lambda) = 0
\end{equation}
into (\ref{SB-SLambda}), we get
\begin{align}
&\lim_{\Lambda \to 0+} e^{- S_\Lambda [\phi] + \frac{1}{2} \int_p
  \frac{p^2}{K(p/\Lambda)} \phi (p) \phi (-p)} = \lim_{\Lambda \to 0+}
  e^{- S_{I,\Lambda} [\phi]} \notag\\
&= e^{- \frac{1}{2} \int_p \frac{p^2}{K (p/\Lambda_0)} \phi (p) \phi
  (-p)} \int \mathcal{D} \varphi\, \exp \left[ - S_B [\varphi] +
  \int_p \frac{p^2}{K(p/\Lambda_0)} \varphi (p) \phi (-p) \right]
\end{align}
Hence, rewriting $\phi (p)$ by $\frac{K(p/\Lambda_0)}{p^2} J (p)$, we
obtain the generating functional of the bare theory as the $\Lambda
\to 0+$ limit of $S_{I,\Lambda}$:
\begin{align}
Z_B [J] &\equiv \int \mathcal{D} \varphi\, \exp \left[ -S_B [\varphi] - \int_p \varphi
  (p) J(-p)\right]\notag\\
&= e^{- \frac{1}{2} \int_pJ(p) J(-p) \frac{K(p/\Lambda_0)}{p^2}}
  \lim_{\Lambda \to 0+} \exp \left( - S_{I,\Lambda} \left[
  \frac{K(p/\Lambda_0)}{p^2} J(p) \right] \right)\label{ZBJ-limit}
\end{align}

\subsubsection{IR limit of a critical theory}

For the bare theory at criticality, we expect that the correlation
functions
\begin{equation}
\vev{\varphi (p_1) \cdots \varphi (p_n)}_B
\equiv \int \mathcal{D} \varphi\, \varphi (p_1) \cdots \varphi
  (p_n)\,e^{-S_B [\varphi]}
\end{equation}
to become scale invariant in the IR limit, i.e., for small momenta.
To be more precise, we can define the limit
\begin{equation}
\mathcal{C} (p_1,\cdots,p_n) \equiv \lim_{t \to \infty} e^{\frac{n}{2}
  \left( -(D+2) + \eta \right) t}\vev{\varphi (p_1 e^{-t}) \cdots
  \varphi (p_n 
  e^{-t})}_B \label{IRlimit}
\end{equation}
where $\frac{\eta}{2}$ is the anomalous dimension.

What does this mean for $S_\Lambda$ in the limit $\Lambda \to 0+$?  As
we have seen above, the interaction part $S_{I,\Lambda}$ becomes the
generating functional of the bare theory in this limit.  Since only
the IR limit of the correlation functions are scale invariant, only
the low momentum part of $\lim_{\Lambda \to 0+} S_{I,\Lambda}$
corresponds to the scale invariant theory defined by the IR limit
(\ref{IRlimit}).

To understand the IR limit better, we follow Wilson \cite{Wilson} and
reformulate the ERG trasnformation in two steps:
\begin{enumerate}
\item introduction of an anomalous dimension (section
  \ref{subsub:anomalous}) --- the anomalous dimension is an important
  ingredient of the IR limit.  We need to introduce an anomalous
  dimension of the field within ERG.
\item introduction of a dimensionless framework (section \ref{subsub:dimensionless}) ---
  each time we lower the cutoff $\Lambda$ we have to rescale space to
  restore the same momentum cutoff.  This is necessary to realize
  scale invariance within ERG.
\end{enumerate}

\subsubsection{Anomalous dimension in ERG\label{subsub:anomalous}}

The cutoff dependent Wilson action $S_\Lambda [\phi]$ has two parts:
\begin{equation}
S_\Lambda [\phi] = \frac{1}{2} \int_p \frac{p^2}{K(p/\Lambda)} \phi
(p) \phi (-p) + S_{I,\Lambda} [\phi]
\end{equation}
The first term is a kinetic term, but this is not the only kinetic
term; part of the interaction quadratic in $\phi$'s also contains the
kinetic term.  The normalization of $\phi$ has no physical
meaning, and it is natural to normalize the field so that
$S_{I,\Lambda}$ contains no kinetic term.

To do this, we modify the ERG differential equation (\ref{ERG-SL}) by
adding a number operator \cite{Igarashi, Igarashi-gamma}:
\begin{align}
- \Lambda \partial_\Lambda S_\Lambda [\phi]
&= \int_p \left( - 2 p^2 \frac{d}{d p^2} \ln K(p/\Lambda)\, \phi (p)
  \frac{\delta S_\Lambda}{\delta \phi (p)} - \frac{d}{dp^2}
  K(p/\Lambda) \lb \frac{\delta^2 S_\Lambda}{\delta \phi (p) \delta
  \phi (-p)} - \frac{\delta S_\Lambda}{\delta \phi (p)} \frac{\delta
  S_\Lambda}{\delta \phi (-p)} \rb \right)\notag\\
&\quad - \frac{\eta_\Lambda}{2} \N_\Lambda [\phi]\label{ERG-SL-eta}
\end{align}
where the number operator $\N_\Lambda [\phi]$ is defined by
\begin{equation}
\N_\Lambda [\phi] \equiv \int_p \left[ \phi (p) \frac{\delta
    S_\Lambda}{\delta \phi (p)} + \frac{K(p/\Lambda) \left(1 -
      K(p/\Lambda)\right)}{p^2} \lb \frac{\delta^2 S_\Lambda}{\delta \phi (p) \delta
  \phi (-p)} - \frac{\delta S_\Lambda}{\delta \phi (p)} \frac{\delta
  S_\Lambda}{\delta \phi (-p)} \rb \right]\label{NL}
\end{equation}
This counts the number of fields:
\begin{equation}
\vev{\N_\Lambda [\phi]\,\phi (p_1) \cdots \phi (p_n)}_{S_\Lambda} = n \vev{\phi
  (p_1) \cdots \phi (p_n)}_{S_\Lambda}
\end{equation}
(Again we are ignoring small corrections to the two-point functions.)
Under (\ref{ERG-SL-eta}) the correlation functions change as
\begin{equation}
\prod_{i=1}^n \frac{1}{K(p_i/\Lambda)} \, \vev{\phi (p_1) \cdots \phi
  (p_n)}_{S_\Lambda} =
\left(\frac{Z_\Lambda}{Z_{\Lambda'}}\right)^{\frac{n}{2}}
\prod_{i=1}^n  \frac{1}{K(p_i/\Lambda')}\, \vev{\phi (p_1) \cdots \phi
  (p_n)}_{S_{\Lambda'}}\label{SL-correlators-eta}
\end{equation}
where $Z_\Lambda$ is the solution of
\begin{equation}
- \Lambda \frac{\partial}{\partial \Lambda} Z_\Lambda = \eta_\Lambda\,
Z_\Lambda
\end{equation}
satisfying the initial condition
\begin{equation}
Z_{\Lambda_0} = 1
\end{equation}
We can choose $\eta_\Lambda$ so that $S_\Lambda$ has the same kinetic
term independent of $\Lambda$.  For (\ref{ERG-SL-eta}), the integral
formula (\ref{SB-SLambda}) must be changed to \cite{Sonoda-equiv}
\begin{align}
e^{S_\Lambda [\phi]}
&= \int \mathcal{D} \varphi\, e^{S_0 [\varphi]}\notag\\
&\quad \times \exp \left[ - \frac{1}{2} \int_p
  \frac{p^2}{\frac{1-K(p/\Lambda)}{Z_\Lambda K(p/\Lambda)} -
  \frac{1-K(p/\Lambda_0)}{K(p/\Lambda_0)}} \left( \frac{\varphi (p)}{K(p/\Lambda_0)} -
  \frac{\phi (p)}{\sqrt{Z_\Lambda}\,K(p/\Lambda)}\right)
\left( \frac{\varphi (-p)}{K(p/\Lambda_0)} -
  \frac{\phi (-p)}{\sqrt{Z_\Lambda}\,K(p/\Lambda)}\right) \right]\label{SB-SLambda-eta}
\end{align}
This reduces to (\ref{SB-SLambda}) for $Z_\Lambda = 1$.

\subsubsection{Dimensionless framework\label{subsub:dimensionless}}

To reach the IR limit (\ref{IRlimit}) we must look at smaller and
smaller momenta as we lower the cutoff $\Lambda$.  We can do this by
measuring the momenta in units of the cutoff $\Lambda$.  At the same
time we render all the dimensionful quantities such as $\phi (p)$
dimensionless by using appropriate powers of $\Lambda$.

We introduce a dimensionless parameter $t$ by
\begin{equation}
\Lambda = \mu \, e^{-t}
\end{equation}
where $\mu$ is an arbitary fixed momentum scale.  We then define the
dimensionless field with dimensionless momentum by
\begin{equation}
\bar{\phi} (p) \equiv \Lambda^{\frac{D+2}{2}} \phi (p \Lambda)
\end{equation}
and define a Wilson action parametrized by $t$:
\begin{equation}
\bar{S}_t [\bar{\phi}] \equiv S_\Lambda [\phi]
\end{equation}
We can now rewrite (\ref{ERG-SL-eta}) for $\bar{S}_t$:
\begin{align}
\partial_t \bar{S}_t [\bar{\phi}]
&= \int_p \left( - 2 p^2 \frac{d}{dp^2} \ln K(p) + p \cdot \partial_p
  + \frac{D+2}{2} \right) \bar{\phi} (p) \cdot \frac{\delta \bar{S}_t
  [\bar{\phi}]}{\delta \bar{\phi} (p)}\notag\\
&\quad + \int_p (-) \frac{d}{dp^2} K(p) \, \lb \frac{\delta^2
  \bar{S}_t}{\delta \bar{\phi} (p) \delta \bar{\phi} (-p)} -
  \frac{\delta \bar{S}_t}{\delta \bar{\phi} (p)}  \frac{\delta
  \bar{S}_t}{\delta \bar{\phi} (-p)} \rb
- \frac{\eta_t}{2} \N_t [\bar{\phi}]
\label{ERG-St-eta}
\end{align}
where we have replaced $\eta_\Lambda$ by $\eta_t$, and
\begin{equation}
\N_t [\bar{\phi}]
\equiv \int_p \bar{\phi} (p) \frac{\delta \bar{S}_t
  [\bar{\phi}]}{\delta \bar{\phi} (p)} + \int_p
\frac{K(p)\left(1-K(p)\right)}{p^2} \left(
  \frac{\delta^2 \bar{S}_t}{\delta \bar{\phi} (p) \delta \bar{\phi}
    (-p)} - \frac{\delta \bar{S}_t}{\delta \bar{\phi} (p)}
  \frac{\delta \bar{S}_t}{\delta \bar{\phi} (-p)} \right)\label{Noperator}
\end{equation}
is the number operator for $\bar{S}_t$.

Rewriting (\ref{SL-correlators-eta}) in terms of dimensionless fields, we
obtain
\begin{align}
&\prod_{i=1}^n \frac{1}{K(p_i)} \,\vev{\bar{\phi} (p_1 ) \cdots \bar{\phi} (p_n)}_{\bar{S}_{t}}\notag\\
&= \left(\frac{Z_t}{Z_{t'}}\right)^{\frac{n}{2}} e^{- \frac{n}{2} \left(D-2\right) (t-t')} \prod_{i=1}^n
\frac{1}{K(p_i e^{-(t-t')})}\, \vev{\bar{\phi} (p_1 e^{-(t-t')}) \cdots \bar{\phi}
  (p_n e^{-(t-t')})}_{\bar{S}_{t'}} \label{St-correlators-eta}
\end{align}
where $Z_t$ satisfies
\begin{equation}
\partial_t Z_t = \eta_t \, Z_t
\end{equation}
(The corrections to the two-point functions are ignored.)  Comparing
(\ref{St-correlators-eta}) with (\ref{IRlimit}), the existence of the
IR limit implies that
\begin{equation}
\lim_{t \to \infty} \eta_t = \eta
\end{equation}
and
\begin{equation}
\lim_{t \to \infty} \prod_{i=1}^n \frac{1}{K(p_i)}\, \vev{\bar{\phi}
  (p_1) \cdots \bar{\phi}}_{\bar{S}_t} = \mathcal{C} (p_1, \cdots,
p_n)
\end{equation}
In other words $\bar{S}_t$ approaches a limit as $t \to +\infty$:
\begin{equation}
\lim_{t \to +\infty} \bar{S}_t = \bar{S}_\infty
\end{equation}
We call $\bar{S}_\infty$ a fixed point because the right-hand side of
(\ref{ERG-St-eta}) vanishes for it:
\begin{align}
0 &= \int_p \left( - 2 p^2 \frac{d}{dp^2} \ln K(p) + p \cdot \partial_p
  + \frac{D+2}{2} \right) \bar{\phi} (p) \cdot \frac{\delta \bar{S}_\infty
  [\bar{\phi}]}{\delta \bar{\phi} (p)}\notag\\
&\quad + \int_p (-) \frac{d}{dp^2} K(p) \, \lb \frac{\delta^2
  \bar{S}_\infty}{\delta \bar{\phi} (p) \delta \bar{\phi} (-p)} -
  \frac{\delta \bar{S}_\infty}{\delta \bar{\phi} (p)}  \frac{\delta
  \bar{S}_\infty}{\delta \bar{\phi} (-p)} \rb - \frac{\eta}{2}
  \N_\infty [\bar{\phi}]
\end{align}

\subsubsection{Fixed-point equation}

Instead of choosing $\eta$ dependent on $t$, we may choose $\eta$ as a
constant so that there is a non-trivial fixed-point solution
$\bar{S}_\infty$ for which the right-hand side of (\ref{ERG-St-eta})
vanishes.  With a constant anomalous dimension, the dimensionless ERG
equation is given by
\begin{align}
\partial_t \bar{S}_t [\bar{\phi}]
&= \int_p \left( - 2 p^2 \frac{d}{dp^2} \ln K(p) + \frac{D+2}{2} -
  \frac{\eta}{2} + p \cdot \partial_p \right) \bar{\phi} (p) \cdot
  \frac{\delta \bar{S}_t [\bar{\phi}]}{\delta \bar{\phi} (p)}\notag\\
&\quad + \int_p \left( - 2 \frac{d}{dp^2} K(p) - \eta \frac{K(p)
  \left(1 - K(p)\right)}{p^2} \right) \frac{1}{2} \left( \frac{\delta^2
  \bar{S}_t [\bar{\phi}]}{\delta \bar{\phi} (p) \delta \bar{\phi}
  (-p)} - \frac{\delta \bar{S}_t [\bar{\phi}]}{\delta \bar{\phi} (p)}
  \frac{\delta \bar{S}_t [\bar{\phi}]}{\delta \bar{\phi} (-p)} \right)
\end{align}
For the O($N$) model with $N$ fields $\phi^i\,(i=1,\cdots,N)$, the ERG
equation becomes
\begin{align}\label{ERG-St-eta-ON}
\partial_t \bar{S}_t [\bar{\phi}]
&= \int_p \left( - 2 p^2 \frac{d}{dp^2} \ln K(p) + \frac{D+2}{2} -
  \frac{\eta}{2} + p \cdot \partial_p \right) \bar{\phi}^i (p) \cdot
  \frac{\delta \bar{S}_t [\bar{\phi}]}{\delta \bar{\phi}^i (p)}\notag\\
&\quad + \int_p \left( - 2 \frac{d}{dp^2} K(p) - \eta \frac{K(p)
  \left(1 - K(p)\right)}{p^2} \right) \frac{1}{2} \left( \frac{\delta^2
  \bar{S}_t [\bar{\phi}]}{\delta \bar{\phi}^i (p) \delta \bar{\phi}^i
  (-p)} - \frac{\delta \bar{S}_t [\bar{\phi}]}{\delta \bar{\phi}^i (p)}
  \frac{\delta \bar{S}_t [\bar{\phi}]}{\delta \bar{\phi}^i (-p)} \right)
\end{align}
where the repeated indices $i$ are summed over.

\subsection{Energy Momentum Tensor: Scale Invariance and Conformal Invariance}
\label{secA2}

\subsubsection{ Energy Momentum Tensor in the Classical Theory}

In this paper we will focus on the following Euclidean action whenever
a concrete action is required for a calculation 
\[
S_E=\int d^Dx \sqrt g [\hf g^{\mu\nu}\p_\mu \phi \p_\nu \phi +\hf m^2\phi^2 +\frac{\la}{4!}\phi^4]
\]
Using 
\[
\dd g =g g^{\mu\nu}\dd g_{\mu\nu},  \quad\dd \sqrt g = \hf \sqrt g
g^{\mu\nu}\dd g_{\mu\nu},\quad \dd g_{\mu\nu}= -g_{\mu\rho}\dd
g^{\rho \sigma}g_{\sigma \nu}
\]
we get
\be
\dd S_E =-\int d^Dx~\hf \dd g_{\mu\nu}\sqrt g [ \p^\mu \phi \p^\nu
\phi - g^{\mu\nu}{\cal L}]\equiv -\int d^Dx~\hf\dd g_{\mu\nu}\sqrt g
T^{\mu\nu} 
\ee
where
\be
T^{\mu\nu}\equiv -\frac{2}{\sqrt g}\frac{\dd S}{\dd g_{\mu\nu}}
= \p^\mu \phi \p^\nu \phi - g^{\mu\nu}{\cal L}
\ee
One can check that
\be\label{cons}
\p^\nu T_{\mu\nu}= -\p_\mu \phi  \left[\frac{\p {\cal L}}{\p \phi}
  -\p^\rho \left(\frac{\p {\cal L}}{\p ^\rho \phi}\right)\right]=-\p_\mu\phi
\frac{\dd S_E}{\dd \phi} 
\ee
Thus, classically the energy momentum tensor is conserved on-shell.
\newline
Now we rewrite $T_{\mu\nu}$ in a form that will be useful later. Define the traceless tensor
\be
t_{\mu\nu}= D \p_\mu\p_\nu - g_{\mu\nu}\Box~~
\ee
and the transverse tensor
\be
\sigma_{\mu\nu}= (g_{\mu\nu}\Box - \p_\mu \p_\nu )\phi^2
\ee
Using the identity
\[
\p_\mu \phi \p_\nu \phi = \p_\mu \p_\nu \hf \phi^2 -\phi \p_\mu\p_\nu \phi
\]
one can rewrite
\begin{align}
\label{classicalem}
 T_{\mu\nu} &=  \frac{1}{4(D-1)} t_{\mu\nu}\phi^2 + \frac{D-2}{4(D-1)}
 (\p_\mu\p_\nu -g_{\mu\nu}\p^2)\phi^2 -\od \phi t_{\mu\nu} \phi \notag\\
&\quad  - \od g_{\mu\nu}  \left[m^2\phi^2 + (4-D)\frac{\la}{4!}\phi^4 +\frac{D-2}{2}E\right]
\end{align}
The trace which is proportional to $g_{\mu\nu}\frac{\dd S}{\dd
  g_{\mu\nu}}$ can be written as $\frac{\partial S}{\partial t}$ when
$g_{\mu\nu}=e^{2t}\dd_{\mu\nu}$ and is the response to scale
transformations. 
\be   \label{Tmn}
T^\mu_\mu= \frac{(2-D)}{4}\Box \phi^2 -
\left[m^2\phi^2+(4-D)\frac{\la}{4!}\phi^4 +\frac{D-2}{2}E\right] 
\ee
with
\[
E =\phi \frac{\dd S_E}{\dd \phi}
\] 
proportional to the equation of motion.  The terms proportional to
$m^2$ and $\la$ are genuine violations of scale invariance. But the
first term can be gotten rid of by defining the improved energy
momentum tensor
\be \label{improvement} \Theta_{\mu\nu}=T_{\mu\nu}+
\frac{D-2}{4(D-1)}\sigma_{\mu\nu}\phi^2 \ee
which is still conserved. So in a genuinely classically scale
invariant theory with $m^2=0$ and $\la=0 ~\mathrm{or}~ D=4$ one expects
\[
\Theta^\mu_\mu= \frac{2-D}{2}E
\]

\subsubsection{Trace of the Energy Momentum Tensor in the Quantum Theory: Perturbative}

When quantum corrections \footnote{We are working in Euclidean
  space. So ``quantum" fluctuations are actually statistical
  fluctuations} are included the condition for scale invariance is
modified. The trace will be defined as before proportional to
$\frac{\partial S}{\partial t}$.  Before we turn to the exact RG let
us see what happens in the usual lowest order perturbation theory.
Let us start at $\lo$ and evolve to $\lm$ with $\lm$ close to $\lo$.
\be S_{\lo} = \int _x \left[\hf \p_\mu\phi \p^\mu \phi + \hf m_0^2
  \phi^2 +\la _0 \frac{\phi^4}{4!}\right] 
\ee
and
\[
S_\lm = \int _x \left[(1-\dd Z(t))\hf \p_\mu\phi \p^\mu \phi + \hf  (m_0^2+
\delta m_0(t)^2)\phi^2 + (\la_0+\dd \la_0(t)) \frac{\phi^4}{4!} +
O(1/\lm)\right] 
\]
Here $\dd Z$ is the correction to the kinetic term coming from the two
loop diagram at $\mathcal{O}(\la^2)$, $\dd m_0^2\approx O(\la)$ and $\dd \la_0
\approx \mathcal{O}(\la^2)$ are the corrections starting at one loop. 
\newline
We rewrite $S_\lm$ in a suggestive way by adding and subtracting some terms proportional to $\dd Z$:
\begin{align}
S_\lm &=
\int _x \Big[\hf \p_\mu\phi \p^\mu \phi + \hf  \underbrace{(m_0^2+
    \delta m_0(t)^2 + \dd Z m_0^2)}_{m^2(t)=m^2_R}\phi^2 +
  \underbrace{(\la_0+\dd \la_0(t)+2\dd Z \la_0)}_{\la(t)=\la_R}
  \frac{\phi'^4}{4!} + \mathcal{O}(1/\lm)\Big] \notag\\
&\quad -\dd Z \underbrace{\left[\hf \p_\mu\phi \p^\mu \phi + \hf
  m_0^2\phi^2+2 \la_0 \frac{\phi^4}{4!}\right]}_{\phi \frac{\p {\cal L}}{\p
  \phi}} 
\end{align}
If we think of $S_{\lo}$ as the bare action $S_B$ and $S_\lm$ as the
renormalized action $S_R$ so that $S_B = S_R + S_{counter-term}$, then
$\la_0=\la_B$ and $\la (t)=\la_R$. The relation between renormalized
and bare quantities is 
\[
\la_B = \frac{\la_R + \dd \la_R}{Z^2}
\] 
Here $\dd \la_R$ is the {\em counterterm} and is chosen to {\em
  cancel} the correction $\dd \la_0$ so $\dd \la_R =- \dd \la_0$.  Let
us write everything in terms of $\la_B$:
\begin{align*}
\la_B 
&= \la_R +\dd \la_R - 2\dd Z \la_R \approx \la_R +\dd \la_R - 2\dd Z \la_0\\
\la_B + 2 \dd Z \la _0 - \dd \la _R 
&= \la_0 + 2 \dd Z \la_0 + \dd \la_0=\la_R=\la(t)
\end{align*}
Thus for small $t$:
\[
\la(t)= \la_0+\beta(\la_0)t~~;~~~m^2(t)=m^2(0)(1+\gamma_m t)~~~;~~\dd Z =-2\gamma t
\]
Furthermore define
\[
x  = \bar x \lm^{-1} = \bar x \lo e^t
\]
The trace of the energy momentum tensor is given by  the dependence on $t$
\begin{align} \label{dtS}
-T^\mu_{~\mu}
&=\frac{\p S_{\lo}}{\p t}\notag\\
&=\lm^{-D}\left\lbrace
\int _{\bar x} \left[ \hf  m_0^2\gamma_m (\la_0)\phi^2 + \beta(\la_0)
  \frac{\phi^4}{4!}\right]+2\gamma \int_x \hf \phi \frac{\dd
  S_{\lo}}{\dd \phi(x)} \right.\notag\\
&\qquad\left.+D \int _{\bar x} [ \hf  m_0^2\phi^2+ \la_0
  \frac{\phi^4}{4!}]+(D-2)\int_{\bar x}\hf \p_{\bar \mu}\phi \p^{\bar
  \mu} \phi  
 + O(1/\lo)]\right\rbrace
\end{align}
Define dimensionless variables as
\[
m_0^2 = \bar m^2 \lo^2 = \bar m^2 e^{2t} \lm^2
\]
and
\[
\la_0 = (\lo)^{4-D}\bar \la_0 = \bar \la_0e^{(4-D)t}(\lm)^{4-D}
\]
and fields
\[
\phi=(\lm)^{\frac{D-2}{2}}\bphi= e^{-\frac{D-2}{2}t}\lo^{\frac{D-2}{2}}\bar \phi
\]
Now add and subtract 
\[
(\frac{D-2}{2}) \int_{\bar x} \bphi \frac{\dd S_{\lo}}{\dd \bphi(x)}
\]
to get
\be \label{dtSdimless} 
-T^\mu_{~\mu}= \int _{\bar x} \underbrace{\Big[ \hf
  \bar m^2(2+\gamma_m (\la_0))\bphi^2 +( \beta(\la_0)-(D-4)\la_0)
  \frac{\bphi^4}{4!}\Big]}_{\textrm{``$\beta$-function''}}+\left(\frac{D-2}{2}+\gamma\right)
\int_{\bar x} \bphi \frac{\dd S_{\lo}}{\dd \bphi(x)} + O(1/\lo) 
\ee
LHS can be identified with the trace of the energy momentum tensor in
the quantum theory and can be compared with the corresponding
classical expression in \eqref{Tmn}. 
The above gives an idea of how the quantum corrections modify $T_{\mu\nu}$.  A detailed calculation of the energy momentum tensor in the renormalized theory in terms of composite operators and using dimensional regularization is given in \cite{Brown}.
  A systematic and precise treatment is provided by ERG and is given in \cite{Sonoda-emt,Rosten2} and is summarized below.

\subsubsection{Energy Momentum Tensor in Exact RG}
\label{emerg}

We summarize the properties of the energy momentum tensor in ERG,  given in \cite{Sonoda-emt}.
\newline
The Ward Identity  almost \footnote{up to transverse terms of the form $\p_\mu\p_\nu-\Box \dd_{\mu\nu}$ that do not contribute} defines the energy momentum. 
Because of general coordinate invariance
\[
\dd x^\mu = -\eps^\mu~~;~~~~\phi'(x) = \phi(x) + \eps^\mu \p_\mu \phi(x)
\]
is equivalent to (Assume that $g_{\mu\nu}=\eta_{\mu\nu}$)
\[
\dd g_{\mu\nu}= \eps_{(\mu,\nu)}
\]
and 
\[
\idp'=\idp_{g+\dd g}~~~;~~~S[\phi,g+\dd g]=S[\phi',g]
\]
Thus the following identity must hold
\[
Z[J]=\idp' e^{-S[\phi'(x)] + \int _{x} J(x)\phi'(x)}=\idp_{g+\dd g}e^{-S[\phi(x),g+\dd g] + \int _xJ(x)(\phi(x) +\eps^\mu \p_\mu \phi(x))}
\]
Then using the definition of the energy momentum tensor, i.e.
\be   \label{defn}
Z[J=0,g+\dd g]=\idp_{g+\dd g} e^{-S[\phi,g+\dd g]}\equiv \idp _g e^{-S[\phi,g]+\hf\int \sqrt g \dd g_{\mu\nu}T^{\mu\nu}}
\ee
we get the Ward identity
\be   \label{wi}
-\p_\mu \langle T^\mu_{~\nu}(x) \phi(x_1)...\phi(x_n)\rangle + \sum_{i=1}^n \dd(x-x_i)\langle \phi(x_1)....\p_\nu\phi(x_i)...\phi(x_n)\rangle=0
\ee
This is a statement of the conservation of $T_{\mu\nu}$ corresponding to the classical statement \eqref{cons}.
\newline
In ERG this can be written as a Ward identity for the composite operator $[T_{\mu\nu}]$
\be    \label{wierg}
q^\mu [T_{\mu\nu}(q)]=\int_p e^{S[\phi]}K(p) (p+q)_\nu\frac{\dd}{\dd \phi(p)}( [\phi(p+q)]e^{-S[\phi]})
\ee
The equation corresponding to \eqref{dtSdimless} and \eqref{Tmn}
is
\be  \label{Tmnerg}
T^\mu_\mu(0)= -\frac{\p S}{\p t} - (\frac{D-2}{2} +\gamma) {\cal N}
\ee
where $-\frac{\p S}{\p t}$ gives the ERG evolution, with anomalous dimension, in terms of dimensioness variables - the ``$\beta$-function". It vanishes at the fixed point. ${\cal N}$ is the number operator. Note that this equation is obtained for zero momentum or as an integral over space-time in position space. The classical analog of this is \eqref{Tmn}, which was obtained for arbitrary momentum. 

Note that in equations \eqref{wierg}  and \eqref{Tmnerg}, both LHS and RHS are composite operators. So one strategy will be to evaluate $T_{\mu\nu}$ using these equations in the bare theory at some scale $\lo$ which will be taken to be infinity. The bare theory is very simple so the calculations can be done exactly. Then one can evolve $T_{\mu\nu}$ down to a scale $\lm<<\lo$ order by order using the ERG evolution operator. If we choose $\la$ and $m$ to be on the critical surface we are guaranteed that at $\lm$ the theory flows to the fixed point action. Thus we will have evaluated the energy momentum tensor at the fixed point.  

Another approach is to work directly with the known fixed point action and solve the Ward identity order by order. In this paper we follow the second approach.

\section{Wilson-Fisher Fixed Point for the $O(N)$ Model}
\label{sec 3}

We will find the fixed-point Wilson action by putting $\frac{\partial \bar S_t}{\partial t}=0$ in (\ref{ERG-St-eta-ON}). As we will work mostly with dimensionless variables  we will remove the bar sign from the dimensionless variables unless otherwise mentioned. Also t dependence of actions and fields being readily implied,  the subscript t will be omitted too.  We give the fixed point action $S$  in the following form:
\[
S= S_2 + S_4 +S_6
\]
where $S_2$ and $S_4$ are given by
\begin{align}
S_2 &= \int \Dp  U_2(p) \hf \phi^I(p) \phi^I(-p) \\
S_4 &= \frac{1}{2}\prod_{i=1}^3\int \frac{d^Dp_i}{(2\pi)^D} U_4 (p _1,
      p _2; p_3,p_4)
      \hf \phi^I(p_1)\phi^I(p_2)\hf \phi^J(p_3)\phi^J(p_4)
\end{align}
where  $p_1+p_2+p_3+p_4=0$ is implied. Instead of putting an explicit delta
function and integrating over $p_4$ we will simply impose momentum
conservation at every stage.  Accordingly $S_6$ is given by
\be
S_6= \frac{1}{3!}\prod_{i=1}^5\int \frac{d^Dp_i}{(2\pi)^D}  U_6 (p _1,
p _2; p_3,p_4;p_5,p_6)
\hf\phi^I(p_1)\phi^I(p_2)
\hf\phi^J(p_3)\phi^J(p_4)\hf\phi^K(p_5)\phi^K(p_6)
\ee

\subsection{Equations for the vertices}
We get the following equations for $U_2$,$U_4$ and $U_6$:
\paragraph{Equation for $U_2$}
\begin{align}\label{V2}
\nonumber 0=\int &\Dp \Bigg\{\bigg(\frac{-\eta}{2} \frac{K(1-K)}{p^2}-K'(p^2)\bigg)\frac{1}{8} \bigg[4N U_4(p_1,-p_1;p,-p)+8 U_4(p_1,p;-p_1,-p)\bigg]\\
 -&\nonumber \frac{1}{2!}2U_2(p)U_2(p)\dd^D(p - p_1)\Bigg\}+\bigg ( \frac{-\eta}{2}+1-2\frac{p_1^2}{K(p_1^2)} K'(p_1^2)\bigg ) U_2(p_1)
  -\frac{1}{2!}p_1\frac{dU_2(p_1)} {dp_1} \\
\end{align}
\paragraph{Equation for $U_4$}
\begin{align}\label{V4}
0=& \nonumber  \int \Dp  \bigg(\frac{-\eta}{2} \frac{K(1-K)}{p^2}-K'(p^2)\bigg)
\frac{1}{48} \\
 \nonumber \times \bigg \lbrace & 6 N U_6 (p _1, p _2;p_3,p_4; p,-p)+ 12 U_6 (p _1,p ;p_2,-p; p_3,p_4)+12 U_6(p_1,p_2;p_3, p;p_4,- p) \bigg \rbrace\\
- \nonumber & \sum_{j=1}^4\bigg(\frac{-\eta}{2} \frac{K(1-K)}{p_j^2}-K'(p_j^2)\bigg)U_2(p_j)~\frac{2}{8}  U_4 (p _1, p _2; p_3,p_4)+\sum_{j=1}^4\bigg(\frac{-\eta}{2}-2\frac{p^2}{K(p_j^2)} K'(p_j^2)\bigg)~\frac{1}{8}  U_4 (p _1, p _2; p_3,p_4)\\
+& \bigg[4-D - \sum_{i=1}^4 p_i \frac{d}{d p_i} \bigg]\frac{1}{8}  U_4 (p _1, p _2; p_3,p_4)
\end{align}
Here $p = p_a+p_b+p_n=-(p_i+p_j+p_m)$.
\paragraph{Equation for $U_6$}
\begin{align}\label{V6}
0=&\nonumber \frac{2}{48} \sum _{6~perm~of~(m,n)}\bigg(\frac{-\eta}{2} \frac{K(1-K)}{(p_i+p_j+p_m)^2)}-K'((p_i+p_j+p_m)^2)\bigg) U_4 (p _i, p _j;p_m,p) U_4 (p _a, p _b;p_n,-p)\\
+ & \nonumber \sum_{j=1}^6\bigg(K'(p_j^2)-\frac{-\eta}{2} \frac{K(1-K)}{p_j^2}\bigg)U_2(p_j)\frac{2}{48}  U_6 (p _1, p _2; p_3,p_4; p_5,p_6) \\
+&  \sum_{j=1}^6\bigg(\frac{-\eta}{2}-2\frac{p^2}{K(p_j^2)} K'(p_j^2)\bigg)\frac{1}{48} U_6 (p _1, p _2; p_3,p_4;p_5,p_6)+\bigg[6 - 2D - \sum_{i=1}^6 p_i \frac{d}{d p_i} \bigg]\frac{1}{48}  U_6 (p _1, p _2; p_3,p_4; p_5,p_6)
\end{align}

\subsection{Solving the Equations}
We know that $U_4 \approx \mathcal{O}(\epsilon)$ and $U_6 \approx \mathcal{O}(\epsilon ^2)$ and $\eta \approx \mathcal{O}(\epsilon^2)$, where $\epsilon = 4-D$.

\subsubsection{$\mathcal{O}(1)$: Retrieving Gaussian theory}
We start with \eqref{V2} for $U_2$. Neglecting $U_4$ and $\eta$ and collecting coefficients of $\phi^2$ we get
\be	
0= K'(p^2)  U_2(p)U_2(p)
+
\bigg(1-2\frac{p^2}{K(p^2)} K'(p^2)\bigg)U_2(p)
  -p^2\frac{dU_2(p)} {dp^2} 
\ee
$U_2(p)= \frac{p^2}{K(p^2)}$ solves this equation. This is expected since the Gaussian theory is expected to be a fixed point - and this ERG was obtained from Polchinski's ERG by adding on the kinetic term
$ \hf \int \Dp \phi(p) \frac{p^2}{K(p^2)}\phi(-p)$.
\vspace{0.1 in}
Thus our solution can be written as
\be \label{U2general}  \boldmath
U_2(p)=  \frac{p^2}{K(p^2)} +\underbrace{U_2^{(1)}(p)}_{\mathcal{O}(\epsilon)}+ \mathcal{O}(\epsilon^2)
\ee

\subsubsection{$\mathcal{O}(\epsilon)$: Fixed Point value of $m^2$}
We go back to \eqref{V2} and keep $U_4$ which is $\mathcal{O}(\epsilon)$ but drop $\eta$ which is $\mathcal{O}(\epsilon^2)$.
\begin{align}
\nonumber 0=\int & \Dp \bigg(\frac{-\eta}{2} \frac{K(1-K)}{p^2}-K'(p^2)\bigg)\times\\
\nonumber \bigg \lbrace & \frac{1}{8}\Big[4N U_4(p_1,p_2;p,-p)+8 U_4(p_1,p;-p,-p_1)\Big] -\frac{1}{2!}2U_2(p)U_2(p)\dd^D(p - p_1)\bigg \rbrace\\
+\nonumber&\bigg(\frac{-\eta}{2}+1-2\frac{p_1^2}{K(p_1^2)} K'(p_1^2)\bigg)U_2(p_1)
  -\frac{1}{2!}p_1\frac{dU_2(p_1)} {dp_1} \\
\end{align}
We use \eqref{U2general} in the above equation and look at the terms of order $\epsilon$. To leading order we set $U_4=\lambda $, which is $\mathcal{O}(\epsilon)$. The equation for $U_2^{(1)}$ is given by
\begin{align*}
0=-\lambda\frac{4N+8}{8} \int \Dp K'(p^2) +2 \frac{p_1^2}{K(p_1^2)} U_2^{(1)}\bigg(p_1)K'(p_1^2)+(1-2\frac{p_1^2}{K(p_1^2)} K'(p_1^2)\bigg)U_2^{(1)}(p_1)
  -p_1^2\frac{dU_2^{(1)}(p_1)} {dp_1^2} 
\end{align*}
To leading order this equation is solved by a constant $U_2^{(1)}$, i.e.
\be 
0= -\lambda\frac{4N+8}{8} \int \Dp K'(p^2) + U_2^{(1)}
\ee
Thus
\be \label{U2epsilon}
\boldsymbol{ U_2^{(1)}=\lambda\frac{N+2}{2} \int \Dp K'(p^2)}
\ee
\vspace{0.1 in}
Here 
\begin{equation*}
\int \Dp = \frac{1}{2^D\pi^{D/2}\Gamma(D/2)}\int (p^2)^{\frac{D-2}{2}}dp^2
\end{equation*}
To get leading results we can set $D=4$:
\be  \label{U2main}
U_2^{(1)}= \lambda\frac{4N+8}{8}\frac{1}{(4\pi)^2}\int _0^\infty dp^2 p^2 K'(p^2)=-\lambda\frac{4N+8}{8}\frac{1}{(4\pi)^2}\int _0^\infty dp^2 K(p^2)
\ee
We have used $K(0)=1, K(\infty)=0$. This gives the fixed point value of the dimensionless mass parameter:
\be	 \label{U2}
U_2^{(1)}=m_\star^2= -\lambda\frac{N+2}{2}\frac{1}{(4\pi)^2}\int _0^\infty dp^2 K(p^2)
\ee
To evaluate the integral explicitly we need a specific form for $K$. We use $K(p^2)=e^{-p^2}$. Then the integral is equal to 1.

\subsubsection {$\mathcal{O}(\epsilon^2)$: Expression for the six-point vertex }
Let us turn to \eqref{V6} reproduced below:
\begin{align}
0=\nonumber -&\frac{2}{48 }\sum _{6~perm~of~(i,j,m)}\bigg(\frac{-\eta}{2} \frac{K(1-K)}{(p_i+p_j+p_m)^2}-K'((p_i+p_j+p_m)^2)\bigg) U_4 (p _i, p _j;p_m,p) U_4 (p _a, p_b;p_n,-p)\\
+& \nonumber \sum_{j=1}^6 \bigg \lbrace\bigg(K'(p_j^2)-\frac{-\eta}{2} \frac{K(1-K)}{p_j^2}\bigg)2 U_2(p_j)+\bigg(\frac{-\eta}{2}-2\frac{p^2}{K(p_j^2)} K'(p_j^2)\bigg) \bigg \rbrace \frac{1}{48} U_6 (p _1, p _2; p_3,p_4;p_5,p_6)\\
+& \bigg[6 - 2D - \sum_{i=1}^6 p_i \frac{d}{d p_i} \bigg]\frac{1}{48}  U_6 (p _1, p _2; p_3,p_4; p_5,p_6)
\end{align}
where $p = p_a+p_b+p_n=-(p_i+p_j+p_m)$.

\vspace{0.05 in}
In this equation we keep terms of $\mathcal{O}(\epsilon^2)$. Since $\eta$ is $\mathcal{O}(\epsilon^2)$, and multiplies terms of $\mathcal{O}(\epsilon^2)$, it contributes only at $\mathcal{O}(\epsilon^4)$ in this equation, so it can be dropped here. Furthermore then, if we use the leading order solution for $U_2= \frac{p^2}{K(p^2)}$, the second and third terms  cancel each other. So we are left with
\begin{align}
0=-&\nonumber\frac{2}{48}\sum _{6~perm~(i,j,m)}K'\big((p_i+p_j+p_m)^2\big) U_4 (p _i, p _j;p_n,p) U_4 (p _a, p _b;p_n,-p)\\
+&\bigg[(6 - 2D - \sum_{i=1}^6 p_i \frac{d}{d p_i}) \bigg]\frac{1}{48}  U_6 (p _1, p _2; p_3,p_4;p_5,p_6)
\end{align}
Since $U_4=\lambda$ to this order, we obtain
\be
0=\lambda^2\frac{2}{48}\sum _{6~perm~(i,j,m)}K'((p_i+p_j+p_m)^2)+\bigg[6 - 2D - \sum_{i=1}^6 p_i \frac{d}{d p_i} \bigg]\frac{1}{48}  U_6 (p _1, p _2; p_3,p_4;p_5,p_6)
\ee
The solution for one permutation is
\[
U_6(p _1,p _2;p_3,p_4;p_5,p _6)=\la^2 \frac{K((p _1+p _2+p _3)^2)-K(0)}{(p_1+p _2+p _3)^2}
\]
The full solution is given by
\begin{align} 
\nonumber U_6(p _1,p _2;p_3,p_4;p_5,p _6)= -\la^2 & \big \lbrace h(p_1+p_2+p_3)+h(p_1+p_2+p_4)+ h(p_1+p_2+p_5)\\
+&h(p_1+p_2+p_6)+h(p_1+p_3+p_4)+h(p_2+p_3+p_4)\big \rbrace
\end{align}
where $h(x)=\frac{K(0)-K(x)}{x^2}$.

\subsubsection{Fixed Point value of $\la$: Solution for $ U_4$ at $\mathcal{O}(\epsilon)$}
The $U_4$ equation is  given by \eqref{V4}. In this equation $\eta$ can be neglected as $-\eta \approx \mathcal{O}(\epsilon^2)$ . Also we put the value of $U_2$ upto order of $\epsilon$ found above. There is a cancellation between the second and third terms on the R.H.S and we obtain
\begin{align} \label{952}
\nonumber &\bigg[\bigg(4-D - \sum_{i=1}^4 p_i \frac{d}{d p_i}\bigg) -\sum_{j=1}^4 2 K'(p_j^2)\frac{\lambda}{16\pi^2}\frac{N+2}{2}\bigg]\frac{1}{8} U_4 (p _1, p _2; p_3,p_4)\\ \nonumber
=&\int \Dp K'(p^2)\frac{1}{48}\bigg \lbrace 6N U_6(p _1, p _2;p_3,p_4; p,-p)+12 U_6 (p _1, p ;p_2,-p; p_3,p_4)+12 U_6(p_1,p_2;p_3,p;p_4,-p)\bigg \rbrace\\
\end{align}
The solution is given in the Appendix \eqref{evaluatev4}. The fixed point value  $\la^*$ given below solves the above equation:
\be \label{lambda}
\boldmath \lambda^* = (4-D) \frac{16\pi^2}{N+8}
\ee

\subsection{Determining Anomalous Dimension}
 
 {\bf $U_2$ equation at $\mathcal{O}(\epsilon^2)$}
\begin{align*}
0 =  \int & \bigg \lbrace \Dp \bigg(\frac{-\eta}{2} \frac{K(1-K)}{p^2}-K'(p^2)\bigg)\left[\frac{\dd^2 S_4}{\dd \phi^I (p) \dd \bphi^I(-p)} -  \frac{\dd S_2}{\dd \phi^I (p)}\frac{\dd S_2}{\dd \phi^I (-p)}\right ] \bigg \rbrace \\
+\bigg \lbrace &-\frac{\eta}{2}-2\frac{p^2}{K(p^2)}K^\prime(p^2)\bigg \rbrace \phi(p).\frac{\delta S}{\delta \phi(p)} +\mathcal{G}_{dil}^c S_2
\end{align*}
where we  plug in:
\br
U_4(p_1,p_2;p_3,p_4)&=& \lambda + \underbrace{\tilde U_4(p_1,p_2;p_3,p_4)}_{O(\epsilon^2)}\nonumber \\
U_2(p)&=& \frac{p^2}{K} -\lambda\frac{N+2}{2} \int \Dp K'(p^2) +  \underbrace{\tilde U_2(p)}_{O(\epsilon^2)}
\er
and keep only $\mathcal{O}(\epsilon^2)$ terms in the above equation to get
\begin{align}
\nonumber 0=\int &\Dp \bigg(\frac{-\eta}{2} \frac{K(1-K)}{p^2}-K'(p^2)\bigg)\times\\
 \nonumber \bigg \lbrace & \frac{1}{8}\bigg[4N\tilde{U}_4(p_1,-p_1;p,-p)+8\tilde{U}_4(p_1,p;-p,-p_1)\bigg] -\frac{1}{2!}2U_2(p)U_2(p)\dd^D(p - p_1)]\bigg \rbrace\\
+&  \bigg(\frac{-\eta}{2}+1-2\frac{p_1^2}{K(p_1^2)} K'(p_1^2)\bigg)U_2(p_1)-p_1^2\frac{dU_2(p_1)} {dp_1^2}
\end{align}

\vspace{0.2 in}
On simplification it gives
\[
-\frac{-\eta}{2} \frac{(1-K)}{K} p_1^2 - \int \Dp K'(p^2)
\frac{1}{8}\bigg[4N\tilde{U}_4(p_1,-p_1;p,-p)+8\tilde{U}_4(p_1,p_1;-p,-p_1)\bigg]+ K^\prime(p_1^2)U_2(p_1)U_2(p_1)\]
\be \label{U2.1}
+ \frac{-\eta}{2} \frac{p_1^2}{K} + \tilde U_2(p_1) -p_1^2\frac{d\tilde U_2(p_1)} {d p_1^2}=0
\ee
In the L.H.S the third term will cancel with part of the second term (shown in \ref{U2cal}). Also the raison d'etre for introducing $\eta$ is to ensure that $U_2 = p^2 +\mathcal{O}(p^4)$. So we let $\tilde U_2 =\mathcal{O}(p^4)$. So The anomalous dimension is given by
\be \label{eta}
\boldmath 
\frac{\eta}{2} = -\frac{d}{dp_1^2} \int \Dp K'(p^2)\frac{1}{8}\bigg[4N\tilde{U}_4^{II}(p_1,-p_1;p,-p)+8\tilde{U}_4^{II}(p_1,p;-p_1,-p)\bigg]~\Bigg|_{p_1^2=0}
\ee
Here the superscript $II$ is explained in Appendix A and refers to a class of Feynman diagrams.
%

\vspace{0.05 in}
$\tilde U_4$ is determined by solving \eqref{952}. So using \eqref{eta} and \eqref{U4.1} one can determine $\eta$. This is done in the Appendix \eqref{ETA}. The result is of course well known \cite{Wilson}:
\be  
\frac{\eta}{2} =\lambda^2\frac{N+2}{4} \frac{1}{(16\pi^2)^2}=\frac{N+2}{(N+8)^2}\frac{\epsilon^2}{4}
\ee

\vspace{0.1 in}

Collecting results we have (we have put D=4 for $\mathcal{O}(\epsilon^2)$ terms),
\begin{align}
\boldsymbol{U_2(p)=  \frac{p^2}{K(p^2)} -\lambda\frac{N+2}{2} \int \Dp K'(p^2) +\tilde U_2(p)}
\end{align}
The expression for $\tilde U_2(p)$ is given in \eqref{tildeU2full} (also in the next section a neater expression is presented).
\begin{align}
 \nonumber \boldsymbol {U_4(p_1,p_2;p_3,p_4) = }& \boldsymbol{(4-D) \frac{16\pi^2}{N+8} +\frac{(N+2)}{2} \frac{\lambda^2}{16\pi^2}\sum_{j=1}^4 h(p_j)}\\
 \boldsymbol{-}& \boldsymbol{\lambda^2 \bigg[(N+4)F(p_1+p_2)+2 F(p_1+p_3)+2 F(p_1+p_4)\bigg]}
\end{align}
where
\begin{align*}
F(p)= \frac{1}{2}\int \Dp h(q)\bigg[h(p+q)-h(q)\bigg]
\end{align*}
and
\begin{align*}
h(p)= \frac{K(0)-K(p^2)}{p^2}
\end{align*}
\begin{boldmath}
\begin{align}
 \nonumber U_6(p _1,p _2;p_3,p_4;p_5,p _6)=-\la^2\bigg \lbrace & h(p_1+p_2+p_3)+h(p_1+p_2+p_4)+h(p_1+p_2+p_5)\\ 
+& h(p_1+p_2+p_6)+h(p_1+p_3+p_4)+h(p_2+p_3+p_4)\bigg \rbrace
\end{align}
\end{boldmath}
and the anomalous dimension is given by
\begin{align} 
 \boldsymbol {\frac{\eta}{2} =\lambda^2\frac{N+2}{4} \frac{1}{(16\pi^2)^2}=\frac{N+2}{(N+8)^2}\frac{\epsilon^2}{4}}
\end{align}
To evaluate the integrals we have put $D=4$ and used specific form of $K(p^2)= e^{-p^2}$.

\vspace{0.1 in}
This completes the solution of the fixed point ERG equation and determination of the eigenvalue $\eta$ corresponding to anomalous dimension up to $O(\eps ^2)$.  In the next section we give a slightly different approach to obtaining the fixed point action and evaluate correlation functions.

\setcounter{subsection}{3}

\section{Correlation functions}

\subsection{A more general equation}

In the previous section we set $\frac{\partial S}{\partial t}=0 $ and solved the fixed point equation for the action order by order.  One can also solve a more general equation  where the LHS is not set to zero but to
$\frac{\partial S}{\partial t}= \beta _J \frac{\partial S}{\p \la _J}$.  The parameters can be chosen so that the beta functions are zero.  This has the effect that the equations are modifed at each order by terms of higher order. The advantage is that the solutions are easier to write down.

\vspace{0.2 in}

We want to obtain the fixed-point Wilson action to order
$\lambda^2$ in the following form:

\begin{align}\label{action}
  S [\phi^I]
  &= \int_p \frac{1}{2} \phi^I (p) \phi^I (-p)\, \left( \frac{p^2}{K(p)} + U_2 (p)
    \right)\nn\\
  &\quad + \frac{1}{2} \int_{p_1,p_2,p_3,p_4} \frac{1}{2} \phi^I (p_1)
    \phi^I (p_2) \frac{1}{2} \phi^J (p_3) \phi^J (p_4)\,\delta
    \left(\sum_{i=1}^4 p_i\right)\, \bigg( \lambda + V_4 (p_1,p_2;
    p_3,p_4)\bigg)\nn\\
  &\quad + \frac{1}{3!} \int_{p_1,\cdots,p_6} \frac{1}{2} \phi^I (p_1)
    \phi^I (p_2) \frac{1}{2} \phi^J (p_3) \phi^J (p_4) \frac{1}{2}
    \phi^K (p_5) \phi^K (p_6) \,\delta
    \left(\sum_{i=1}^6 p_i\right)\\
  &\qquad\qquad \times V_6 (p_1,p_2; p_3,p_4; p_5,p_6)\nn
\end{align}

As we will all vertex function in powers of
 $\lambda$ we have to put the general expression for $\frac{\partial \lambda}{\partial t}$ i.e

\begin{align*}
\frac{\partial \lambda}{\partial t}= (\epsilon \lambda+ \beta_{N}^{(1)}\lambda^2)
\end{align*}

Where $\beta_N^{(1)}$ , the leading term in the beta function, is given by

\begin{align*}
\beta_N^{(1)}=2(N+8) \int \Dp K^\prime(p) \frac{K((0)-K(p)}{p^2}\equiv -(N+8) \int_p f(p)h(p)
\end{align*} 
where $f(p)=-2K^\prime(p^2)$.

If we assume $ V_2(p)= \lambda v_2^{(1)}(p)+ \lambda^2 v_2^{(2)}(p)=  \lambda v_2^{(1)}(p)+ \Big( V_2^I(p)+  V_2^{II}(p)\Big)$, where $V_2^{I(II)}$ is analog of $\tilde U_2^{I(II)}$ in \ref{U2cal}, then

\begin{align*}
\frac{\partial V_2(p)}{\partial t}= \Big(\epsilon \lambda+ \beta_{N}^{(1)}\lambda^2\Big)v_2^{(1)}(p)+ 2\lambda^2 \epsilon v_2^{(2)}(p)+ 2\lambda^3\beta_N^{(1)}v_2^{(2)}(p)
\end{align*}

Similarly if $ V_4(p_1,p_2;p_3,p_4)= V_4^I(p_1,p_2;p_3,p_4)+  V_4^{II}(p_1,p_2;p_3,p_4)$, where $ V_4^{I(II)}(p_1,p_2;p_3,p_4)$ is equivalent to $\tilde U_4^{I(II)}(p_1,p_2;p_3,p_4)$ in \ref{U4cal}.

\begin{align*}
\frac{\partial}{\partial t} \Big[\lambda+V_4(p_1,p_2;p_3,p_4)\Big]= \Big(\epsilon \lambda+ \beta_{N}^{(1)}\lambda^2\Big)+ 2V_4(p_1,p_2;p_3,p_4)\Big(\epsilon+ \beta_{N}^{(1)}\lambda\Big)
\end{align*}

\vspace{0.2 in}

\textbf{A}.  \eqref{U2epsilon} is modified to
\begin{align*}
\frac{1}{2}\epsilon~ v_2^{(1)}(p)=- \frac{4N+8}{8}\int \Dp K^\prime(p^2)+v_2^{(1)}(p).
\end{align*}
gives
\begin{align}
v_2^{(1)}(p)= &- \frac{N+2}{2-\epsilon}\frac{1}{2} \int \Dp f(p)\\
\equiv \nonumber & -(N+2)v_2
\end{align}
where  $v_2= \frac{1}{2-\epsilon}\frac{1}{2} \int \Dp f(p)$

\vspace{0.2 in}

\textbf{B}. \eqref{U4tilde} turns into
\begin{align}
 & \nonumber \bigg[ \epsilon + \sum_{j=1}^4p_j\frac{d}{dp_j} \bigg ] V_4^{II}(p_1,p_2;p_3,p_4))\\
=& -2\lambda^2\int _{\bp} K'(p^2)\bigg[(N+4)h(p_1+p_2+p)+2 h(p+p_1+p_3)+2 h(p+p_1+p_4)-(N+8)h(p)\bigg]
\end{align}
If we write $V_4^{II}(p_1,p_2;p_3,p_4))=-\lambda^2 \Big\lbrace (N+4)F(p_1+p_2)+2F(p_1+p_3)+2F(p_1+p_4)\Big\rbrace$ the equation for $F(p)$ can be written as,
\begin{align}
\big(p.\partial p+\epsilon \big)F(p)=\int \Dp f(q)h(q+p)+\frac{1}{3}\beta^{(1)}
\end{align}
where 
\begin{align*}
\frac{1}{3}\beta^{(1)}= -\int \Dp f(p)h(p)
\end{align*}
The solution , analytic at $p=0$ is,
\begin{align}
F(q)= \frac{1}{2} \int \Dp h(p)\Big(h(q+p)-h(p)\Big)
\end{align}

\vspace{0.2 in}
\textbf{C}. Similarly \eqref{U4I} gets modified to,
\begin{align}
\bigg[\epsilon+ \sum_{j=1}^4 p_j\frac{d}{d p_j}\bigg] \frac{1}{8} V_4^I(p_1,p_2;p_3,p_4)= \lambda^2 (N+2) \int \Dp K^\prime(p^2) \bigg \lbrace  -\frac{1}{8}\sum_{j=1}^4 h(p_j)- \frac{1}{4(2-\epsilon)}   K^\prime(p_j^2)\bigg \rbrace
\end{align}
whose solution is,
\begin{align}
V_4^I(p_1,p_2;p_3,p_4)&= \lambda^2\frac{(N+2)}{2-\epsilon}\int \Dp(- K^\prime(p^2)) \sum_{j=1}^4 h(p_j)
\end{align}
Also
\begin{align*}
\frac{1}{8} \Big \lbrace 4N V_4^I(p_1,-p_1;p,-p)+8 V_4^I(p,p_1;-p,-p_1) \Big\rbrace &= \frac{(N+2)^2}{2-\epsilon} \lambda^2 \int \frac{d^D q}{(2\pi)^D}(-K^\prime( q^2)) \Big[ h(p_1)+h(p) \Big]
\end{align*}

\vspace{0.2 in}
\textbf{D}. \eqref{U2.1} turns into,
\begin{align}
(2-2\epsilon)V_2^I-2p_1^2\frac{d V_2^I(p_1)}{dp_1^2}=~&-\frac{2\lambda^2}{2-\epsilon} (N+2)^2 \bigg \lbrace\int \Dp (-K^\prime(p^2)) \bigg \rbrace^2 h(p_1)-2\big(v_2^{(1)}\big)^2 K^\prime(p_1^2)
\end{align}
The solution is
\begin{align}
\nonumber V_2^I(p_1)= -(N+2)^2\lambda^2 \frac{1}{(2-\epsilon)^2} \frac{1}{4}\bigg \lbrace \int \Dp f(p) \bigg \rbrace^2 h(p_1)
\end{align}

\vspace{0.2 in}
\textbf{E}. \eqref{tildeU2.2} changes to
\begin{align*}
& \nonumber \Big(-2+2\epsilon\Big)V_2^{II}(p_1)+\beta_{N}^{(1)}\lambda^2 v_2^{(1)}(p)+2p_1^2\frac{dV_2^{II}(p_1)} {dp_1^2}\\
=&  -3\lambda^2(N+2)\int_{ r, p}(- K'( p^2))h( r)
\Big[h(p_1 + p+ r)- h( r)\Big]+\frac{2}{2-\epsilon}\Big[(N+2)^2\lambda^2 \int (-K^\prime( q^2))\Big]\int_{p}(-K^\prime(p^2))h(p)-\eta ~p_1^2
\end{align*}
If we assume
\begin{align*}
V_2^{II}(p)= -3\lambda^2 (N+2) G(p)
\end{align*}
Then $G(p)$  satisfies the following equation,
\begin{equation}
(p.\partial p-2+2\epsilon)G(p)= \int f(q)F(p+q)+\frac{2v_2}{3}\int_p f(p) h(p) +\eta^{(2)} p^2
\end{equation}
From \eqref{eta} we get $\eta= 3(N+2) \lambda^2 \eta^{(2)}$ where,
\begin{align*}
\eta^{(2)}= -\frac{d}{dp^2}\int f(q)F(q+p)\Big |_{p=0}
\end{align*}
The solution , analytic at $p=0$ is
\begin{align}
G(p)=  \frac{1}{3} \int h(q)(F(p+q)-F(q))+\frac{1}{\epsilon} \frac{\eta^{(2)}}{2}p^2-\frac{1}{2-2\epsilon}\Bigg(\int f(q)F(q)+\frac{2v_2}{3}\int_p f(p) h(p)  \Bigg)
\end{align}
$V_2^I(p)+V_2^{II}(p)$ when calculated in the limit $\epsilon \rightarrow 0$ gives the expression of $\tilde{U}_2(p)$ mentioned in the previous section.

\vspace{0.2 in}
The solutions are given by,

\begin{subequations}
  \begin{align}
    \boldsymbol { V_2 (p)}
    & \boldsymbol {= - \lambda (N+2) v_2 - \lambda^2 \left( 3 (N+2) G (p) +
      (N+2)^2 \left(v_2\right)^2 h (p) \right)}\\
     \boldsymbol {V_4 (p_1,p_2; p_3,p_4)}
    & \boldsymbol {= - \lambda^2 \Big( (N+4) F(p_1+p_2) + 2 F(p_1+p_3) + 2
      F(p_1+p_4)} \nn\\
    &\qquad\qquad  \boldsymbol { - (N+2) v_2 \sum_{i=1}^4 h(p_i)} \Big)\\
     \boldsymbol {V_6 (p_1,p_2; p_3,p_4; p_5,p_6)}
    & \boldsymbol { = - \lambda^2}     \left( \boldsymbol  {h(p_1+p_2+p_3) + h(p_1+p_2+p_4) +
      h(p_1+p_2+p_5)} \right.\nn\\
    &\qquad\left.  \boldsymbol { + h(p_1+p_2+p_6) + h(p_3+p_4+p_1) +
      h(p_3+p_4+p_2)} \right)
  \end{align}
\end{subequations}
where
\begin{equation*}
f(p)=-2K^\prime(p^2);~h(p)= \frac{K(0)-K(p^2)}{p^2}
\end{equation*}
and
\begin{equation}
 \boldsymbol {  v_2 = \frac{1}{2-\ep} \frac{1}{2} \int \Dp f(p) }
  \label{summary-v2}
\end{equation}
If we take the limit $\epsilon \rightarrow 0$ and $K(p^2)=e^{-p^2}$ we get
\begin{align*}
v_2= \frac{1}{2}\int \frac{d^4 p}{(2\pi)^4} e^{-p^2} = \frac{1}{2}\frac{1}{16\pi^2}
\end{align*}
\begin{equation*}
F(p)= \frac{1}{2}\int \Dp h(q)\Big[h(p+q)-h(q)\Big]
\end{equation*}
The coupling constant $\lambda$ is given, to order $\ep = 4-D$, as
\begin{equation}
   \boldsymbol {\lambda = \frac{\ep}{- \beta_N^{(1)}} = \frac{(4
    \pi)^2}{N+8}}\,\ep \label{summary-lambda} 
\end{equation}
The anomalous dimension is given, to order $\ep^2$, as
\begin{equation}
 \boldsymbol {\eta = \frac{N+2}{2 (N+8)^2}}\,
 \boldsymbol {\ep^2} \label{summary-eta} 
\end{equation}


\subsection{Calculation of correlation functions}

In this section we will calculate two-, four-, and six-point correlation functions.  Recall that our Wilson action has a fixed momentum cutoff
of order $1$.  If we consider the momenta much larger than the cutoff,
the vertices of the Wilson action gives the correlation functions \cite{Sonoda-correlation}. We first rescale the
field
\begin{equation}
J^I (p) \equiv \frac{1}{h(p)} \phi^I (p)
\end{equation}
and define
\begin{equation}
W [J^I] \equiv - S [\phi^I] + \frac{1}{2} \int_p J^I (p) J^I (-p)
\frac{h(p)}{K(p)}
\end{equation}
For our Wilson action, this is given by
\begin{align}
W[J^I]
&= \int_p \frac{1}{2} J^I (p) J^I (-p) \, h(p)^2 \left( \frac{1}{h(p)} - V_2 (p)
  \right)\nn\\
&\quad + \frac{1}{2} \int_{p_1,p_2,p_3,p_4} \frac{1}{2} J^I (p_1) J^I
  (p_2) \frac{1}{2} J^J (p_3) J^J (p_4)\, \delta \left(\sum_{i=1}^4
  p_i\right)\nn\\
&\qquad\qquad \times  \prod_{i=1}^4 h(p_i)\, \cdot \left( - \lambda -
  V_4 (p_1,p_2; p_3,p_4) \right)\nn\\ 
&\quad + \frac{1}{3!} \int_{p_1,\cdots,p_6} \frac{1}{2} J^I (p_1) J^I
  (p_2) \frac{1}{2} J^J (p_3) J^J (p_4) \frac{1}{2} J^K (p_5) J^K
  (p_6)\, \delta \left(\sum_{i=1}^6  p_i\right)\nn\\
&\qquad\qquad \times  \prod_{i=1}^6 h(p_i)\, \cdot (-) V_6 (p_1, p_2;
  p_3,p_4; p_5,p_6)
\end{align}

In the high momentum limit we obtain the generating functional of the
connected correlation functions
\begin{equation}
\mathcal{W} [J^I] = \lim_{t \to +\infty} W [J_t^I]
\end{equation}
where
\begin{equation}
J_t^I (p) \equiv \exp \left( - t \frac{D-2+\eta}{2} \right) J^I (p e^{-t})
\end{equation}
In our case we obtain
\begin{align}
  W [J_t^I] 
  &=\int_p \frac{1}{2} J^I (p)
    J^I (-p) \,  \exp \left( t (2-\eta) \right) h(p e^t)^2 \left(
    \frac{1}{h (p e^{t})} - V_2 (p e^t) 
    \right)\nn\\
  &\quad + \frac{1}{2}
    \int_{p_1,p_2,p_3,p_4} \frac{1}{2} J^I (p_1) J^I 
    (p_2) \frac{1}{2} J^J (p_3) J^J (p_4)\, \delta \left(\sum_{i=1}^4
    p_i\right)\nn\\
  &\qquad\quad \times   \exp \left( t (D+4-2\eta \right) \prod_{i=1}^4
    h(p_i e^t)\, \cdot \left( - 
    \lambda - V_4 (p_1 e^t,p_2 e^t; p_3 e^t,p_4 e^t) \right)\nn\\
  &\quad +  \frac{1}{3!}
    \int_{p_1,\cdots,p_6} \frac{1}{2} J^I (p_1) J^I 
    (p_2) \frac{1}{2} J^J (p_3) J^J (p_4) \frac{1}{2} J^K (p_5) J^K
    (p_6)\, \delta \left(\sum_{i=1}^6  p_i\right)\nn\\
  &\qquad\quad \times   \exp \left( t (2D + 6 - 3 \eta\right)
    \prod_{i=1}^6 h(p_i e^t)\, \cdot (-) V_6 (p_1 e^t, p_2 e^t; p_3
    e^t,p_4 e^t; p_5 e^t,p_6 e^t) 
\end{align}
In the limit $t \to +\infty$ we obtain
\begin{align}
\mathcal{W} [J^I]
&= \int_p \frac{1}{2} J^I (p) J^I (-p)\, C_2 (p)\nn\\
&\quad + \frac{1}{2} \int_{p_1,p_2,p_3,p_4} \frac{1}{2} J^I (p_1) J^I
  (p_2) \frac{1}{2} J^J (p_3) J^J (p_4)\, \delta \left( \sum_{i=1}^4
  p_i \right) \, C_4 (p_1, p_2; p_3, p_4)\nn\\
&\quad + \frac{1}{3!}  \int_{p_1,\cdots, p_6} \frac{1}{2} J^I (p_1) J^I
  (p_2) \frac{1}{2} J^J (p_3) J^J (p_4) \frac{1}{2} J^K (p_5) J^K
  (p_6) \, \delta \left( \sum_{i=1}^6 p_i \right)\nn\\
&\qquad\qquad \times C_6 (p_1, p_2; p_3, p_4; p_5, p_6)
\end{align}

\subsubsection{Two-point function}

\begin{align}
C_2 (p)
&= \lim_{t \to +\infty} \exp \left( t (2-\eta)\right) h(p e^t)^2
  \left( \frac{1}{h(p e^{t})} - V_2 (p e^t) \right)\nn\\
&= \lim_{t \to +\infty}  \frac{1}{(p^2)^2}\left[p^2 (1 - \eta \,t) +
  \lambda^2 3 (N+2) e^{-2t} G (p e^t) \right]
\end{align}
Using
\begin{equation}
G (p e^t) \overset{t \to \infty}{\longrightarrow} p^2 e^{2t}
\frac{1}{12 (4\pi)^4}  \ln \left( p^2 e^{2t}\right)
\end{equation}
we obtain
\begin{equation}
\boxed{
C_2 (p) = \frac{1}{p^2} \left( 1 + \frac{\eta}{2} \ln p^2 \right) =
\frac{1}{p^{2 - \eta}}}
\end{equation}

\subsubsection{Four-point function}

\begin{align}
& C_4 (p_1, p_2; p_3,p_4)\nn\\
&= \lim_{t \to +\infty} \exp \left( t (D+4-2\eta) \right)
  \prod_{i=1}^4 h(p_i e^t) \, \cdot \left( - \lambda - V_4 (p_1 e^t,
  p_2 e^t; p_3 e^t, p_4 e^t ) \right)\nn\\
&= \prod_{i=1}^4 \frac{1}{p_i^2} \lim_{t \to +\infty}  \left(1 - \ep\,
  t \right) \Big[  - \lambda \nn\\
&\qquad + \lambda^2 \left( (N+4) F \left(
  (p_1+p_2) e^t \right) + 2 F \left( (p_1+p_3) e^t \right) + 2 F
  \left( (p_2+p_3) e^t \right) \right)\Big]
\end{align}
Using
\begin{equation}
F (p e^t) \overset{t \to +\infty}{\longrightarrow}- \frac{1}{(4 \pi)^2}
\ln \left( p e^t \right)
\end{equation}
we obtain
\be
\prod_{i=1}^4 p_i^2 \cdot C_4 (p_1,p_2; p_3,p_4)
= - \lambda \left( 1 + \ep \frac{1}{N+8} \ln
  \lb (p_1+p_2)^{N+4} (p_1+p_3)^2 (p_2+p_3)^2 \rb \right)
\ee


\subsubsection{Six-point function}

Since $V_6$ is already of order $\lambda^2$, we can take $D=4$ and
$\eta=0$ to obtain
\begin{align}
C_6 (p_1,p_2;p_3,p_4;p_5,p_6) 
&= \lim_{t\to +\infty} e^{t \left(2 D + 6 - 3 \eta\right)}
  \prod_{i=1}^6 h(p_i e^t)\, (-) V_6 (p_1 e^t, p_2 e^t; p_3 e^t, p_4
  e^t; p_5 e^t, p_6 e^t)\notag\\
& = \lim_{t \to +\infty} e^{14 t}
                                \prod_{i=1}^6 \frac{1}{p_i^2 e^{2t}} 
\cdot \lambda^2 \left( h \left( (p_1+p_2+p_3) e^t \right) + \cdots
  \right)\nn\\
&= \lambda^2 \prod_{i=1}^6 \frac{1}{p_i^2} \left(
  \frac{1}{(p_1+p_2+p_3)^2} + \cdots + \frac{1}{(p_3+p_4+p_2)^2}
  \right)
\end{align}

\section{Construction of the energy-momentum tensor at the fixed
  point}

Given a fixed-point Wilson action, we wish to construct the
energy-momentum tensor $\Theta_{\mu\nu} (p)$.  It is a
\textbf{symmetric} tensor implicitly determined by the Ward identity
\begin{equation}
  p_\mu \Theta_{\mu\nu} (p) = e^{S} \int_q K(q) (q+p)_\nu
  \frac{\delta}{\delta \phi^I (q)} \left( \op{\phi^I (q+p)}\,e^{-S}
  \right) \label{Ward-EM}
\end{equation}
where
\begin{equation}
  \op{\phi^I (p)} \equiv \frac{1}{K(p)} \left(\phi^I (p) -
    \frac{K(p)\left(1-K(p)\right)}{p^2} \frac{\delta S}{\delta \phi^I
      (-p)} \right)
\end{equation}
is the composite operator corresponding to $\phi^I (p)$.  The Ward
identity leaves an additive ambiguity of the form
\[
  \left(p^2 \delta_{\mu\nu} - p_\mu p_\nu \right) \Op (p)
\]
where $\Op (p)$ is a scalar composite operator.  Since
$\Theta_{\mu\nu}$ must have zero scale dimension, $\Op$ must have
scale dimension $-2$.  There is no such $\Op$, since the squared mass
operator $\frac{1}{2} \phi^2$ acquires a positive anomalous dimension
at the fixed point.  Hence, the Ward identity determines
$\Theta_{\mu\nu}$ unambiguously.  In fact we are going to calculate
$\Theta_{\mu\nu} (p)$ only at $p=0$; we need not worry about this
ambiguity anyway.

It is convenient to expand $\Theta_{\mu\nu} (p)$ in powers of $\op{\phi^I}$:
\begin{align}
  \Theta_{\mu\nu} (p)
  &= \sum_{n=0}^\infty \frac{1}{n!}
  \int_{p_1,\cdots,p_{2n}} \prod_{i=1}^n \frac{1}{2} \op{\phi^{I_i}
  (p_{2i-1})} \op{\phi^{I_i} (p_{2i})}\, \delta \left(\sum_{i=1}^{2n} p_i -
    p\right)\nn\\
  &\quad \times c_{\mu\nu, 2n} (p_1,p_2; \cdots; p_{2n-1}, p_{2n})
\end{align}
To order $\lambda^2$, we only have three coefficients
$c_{\mu\nu,0}, c_{\mu\nu, 2}, c_{\mu\nu,4}$.  Since the
field-independent term ($n=0$) is proportional to $\delta (p)$, we
cannot determine $c_{\mu\nu,0}$ from the Ward identity.  So, we will
determine only $c_{\mu\nu,2}$ and $c_{\mu\nu,4}$.

From (\ref{action}), we obtain
\begin{align}
&\op{\phi^I (p)}
= \phi^I (p) - h(p) \Big\lbrace V_2 (p) \phi^I (p) \nn\\
&\quad + \int_{p_1,p_2,p_3} \frac{1}{2} \phi^J (p_1) \phi^J (p_2)
  \phi^I (p_3) \, \delta \left(\sum_{i=1}^3 p_i - p \right)\, \left(
  \lambda + V_4(p_1, p_2; p_3, -p) \right)\nn\\
&\quad + \frac{1}{2} \int_{p_1,\cdots,p_5} \frac{1}{2} \phi^J (p_1)
  \phi^J (p_2) \frac{1}{2} \phi^K (p_3) \phi^K (p_4) \phi^I
  (p_5)\,\delta \left(\sum_{i=1}^5 p_i - p \right)\nn\\
&\qquad\qquad\qquad \times V_6 (p_1,p_2; p_3,p_4; p_5,-p) \Big\rbrace
\end{align}
Inverting this we obtain, to order $\lambda^2$,
\begin{align}
&\phi^I (p)
= \op{\phi^I (p)} + h (p) \Big\lbrace V_2^{1PI} (p) \op{\phi^I (p)} \nn\\
&\quad + \int_{p_1,p_2,p_3} \frac{1}{2} \op{\phi^J (p_1)}\op{\phi^J (p_2)}
  \op{\phi^I (p_3)}\,\delta \left(\sum_{i=1}^3 p_i - p \right) \,
  \left( \lambda + V_4^{1PI}
  (p_1, p_2; p_3, -p)\right) \Big\rbrace
\end{align}
where we have defined the 1PI vertices as
\begin{subequations}
\begin{align}
V_2^{1PI} (p)
&= - \lambda (N+2) v_2 - \lambda^2 3 (N+2) G(p)\\
V_4^{1PI} (p_1, p_2; p_3,p_4)
&= - \lambda^2 \left( (N+4) F(p_1+p_2) + 2 F(p_1+p_3) + 2
  F(p_1+p_4)\right)
\end{align}
\end{subequations}
Note that $\phi^I$ has no sixth order term expanded in $\op{\phi}$'s
to order $\lambda^2$.

The rhs of (\ref{Ward-EM}) gives
\begin{align}
  & e^S \int_q K(q) (q+p)_\nu \frac{\delta}{\delta \phi^I (q)}
    \left( \op{\phi^I (q+p)} e^{-S} \right)\nn\\
  &= \int_q K(q) (q+p)_\nu \left( - \op{\phi^I (q+p)} \frac{\delta
    S}{\delta \phi^I (q)} + \frac{\delta}{\delta \phi^I (q)}
    \op{\phi^I (q+p)} \right)\label{rhs-Ward-EM}
\end{align}
Expanding this in powers of $\op{\phi}$'s, we obtain from
(\ref{Ward-EM}) the following equations that determine the
coefficients $c_{\mu\nu, 2}$ and $c_{\mu\nu, 4}$.
\begin{subequations}
\begin{align}
& p_\mu c_{\mu\nu,2} (p_1, p_2)
  = - p_{1\nu} p_2^2 - p_{2\nu} p_1^2 \nn\\
  &\quad + \lambda (N+2) \left(v_2 p_\nu -
  \int_q (q+p)_\nu R(q) h(q) h(q+p)\right)\nn\\
  &\quad + \lambda^2 (N+2) \Big[ 3 \left( p_{1\nu} G(p_2) + p_{2\nu}
    G(p_1) \right) \nn\\
  &\qquad - (N+2) v_2 \int_q (q+p)_\nu R(q) h(q) h(q+p)
    \left(h(q)+h(q+p)\right) \nn\\
  &\qquad + \frac{1}{2} \int_q \lb (q+p)_\nu R(q) - q_\nu R(q+p) \rb
    h(q) h(q+p) \nn\\
  &\qquad\qquad\quad\times \lb (N+2) F (p) + 3 F(q+p_1) + 3 F(q+p_2) \rb \Big]
    \label{pc2}
\end{align}
and
\begin{align}
&p_\mu c_{\mu\nu,4} (p_1, p_2; p_3,p_4)
  = - \lambda p_\nu \nn\\
  &\quad + \lambda^2 \Big\lbrace (N+4) \left( F(p_1+p_2) (p_3+p_4)_\nu
    + F(p_3+p_4) (p_1+p_2)_\nu \right)\nn\\
  &\qquad+ 2 p_{1\nu} \left(F(p_2+p_3) + F(p_2+p_4)\right) + 2 p_{2\nu}
    \left( F(p_2+p_3) + F(p_2+p_4) \right)\nn\\
  &\qquad+ 2 p_{3\nu} \left(F(p_4+p_1) + F(p_4+p_2) \right) + 2
    p_{4\nu} \left( F(p_3+p_1) + F(p_3+p_2) \right) \Big\rbrace\nn\\
  &\quad + \lambda^2 \frac{1}{2} \int_q \lb (q+p)_\nu R(q) - q_\nu R(q+p)
    \rb h(q) h(q+p)\nn\\
  &\,\times \lb (N+4) \left( h(q+p_1+p_2) + h(q+p_3+p_4) \right) +
    4 \left( h(q+p_1+p_3) + h(q+p_1+p_4) \right)\rb
    \label{pc4}
\end{align}
\end{subequations}
To determine $c_{\mu\nu,2} (p_1, p_2)$ at $p=0$, we substitute
$p_2 = p-p_1$ into the rhs of (\ref{pc2}), and expand the result to
first order in $p$.  This gives
\begin{align}
c_{\mu\nu,2} (p_1, -p_1)
&= - p_1^2 \delta_{\mu\nu} + 2 p_{1\mu} p_{1\nu}\nn\\
&\quad + \lambda (N+2) \delta_{\mu\nu} \lb v_2 - \int_q R(q) \left(
  h(q)^2 + \frac{1}{D} h(q) q \cdot \partial_q h(q) \right)\rb\nn\\
&\quad + \lambda^2 (N+2) \Big\lbrace 3 \left( \delta_{\mu\nu} G(p_1) - 2
  p_{1\mu} p_{1\nu} G' (p_1) \right)\nn\\
&\qquad + \int_q \left( \delta_{\mu\nu} R(q) - 2 q_\mu q_\nu
  R' (q) \right) h(q)^2 \left( - (N+2) v_2 h(q) + 3 F(q+p_1)\right)
  \Big\rbrace \label{cmunu2}
\end{align}
Similarly, substituting $p_4 = p - (p_1+p_2+p_3)$ into the rhs of
(\ref{pc4}) and expanding the result to first order in $p$, we obtain
\begin{align}
&c_{\mu\nu,4} (p_1,p_2;p_3, - (p_1+p_2+p_3)) = - \lambda
  \delta_{\mu\nu}\nn\\
&\quad + \lambda^2 \Big\lbrace 
(N+4) \left( \delta_{\mu\nu} F(p_1+p_2) - 2 (p_1+p_2)_\mu
  (p_1+p_2)_\nu F' (p_1+p_2) \right)\nn\\
&\qquad + 2 \left( \delta_{\mu\nu} F(p_1+p_3) - 2 (p_1+p_3)_\mu
  (p_1+p_3)_\nu F' (p_1+p_3) \right)\nn\\
&\qquad + 2 \left( \delta_{\mu\nu} F(p_2+p_3) - 2 (p_2+p_3)_\mu
  (p_2+p_3)_\nu F' (p_2+p_3) \right)\nn\\
&\qquad + \int_q \left( \delta_{\mu\nu} R(q) - 2 q_\mu q_\nu R' (q)
  \right) h(q)^2 \nn\\
&\qquad \quad \times \left( (N+4) h(q+p_1+p_2) + 2 h(q+p_1+p_3) + 2
  h(q+p_2+p_3)\right) \Big\rbrace\label{cmunu4}
\end{align}

\subsection*{Check of the trace anomaly}

Using the energy-momentum tensor obtained above, we can verify the
trace anomaly
\begin{equation}
\Theta (0) = - \left( \frac{D-2}{2} + \frac{1}{2} \eta \right) \N (0)
\end{equation}
where the anomalous dimension is given by (\ref{summary-eta}) to order
$\ep^2$.

The trace is easily obtained from (\ref{cmunu2}, \ref{cmunu4}) as
\begin{align}
\Theta (0)
&= \int_p \frac{1}{2} \op{\phi^I (p)} \op{\phi^I (-p)} \Bigg[ - (D-2)
  p^2 \nn\\
&\quad + \lambda (N+2) D \lb v_2 - \int_q R(q) \left( h(q)^2 +
  \frac{1}{D} h(q) q\cdot\partial_q h(q) \right)\rb\nn\\
&\quad + \lambda^2 (N+2) \Big\lbrace
3 (D - p \cdot \partial_p ) G (p)\nn\\
&\qquad + \int_q \left(D - q \cdot \partial_q\right) R(q)
  \cdot h(q)^2 \left( - (N+2) v_2 + 3 F(q+p) \right)
  \Big\rbrace\Bigg]\nn\\
&\quad + \frac{1}{2} \int_{p_1,\cdots,p_4} \frac{1}{2} \op{\phi^I
  (p_1)} \op{\phi^I (p_2)} \frac{1}{2} \op{\phi^J (p_3)} \op{\phi^J
  (p_4)}\, \delta \left(\sum_{i=1}^4 p_i\right)\nn\\
&\qquad \times \Bigg[ - \lambda D\nn\\
&\qquad\quad + \lambda^2 \Big\lbrace
  (N+4) \left(D - p \cdot \partial_p \right)
  F(p)\Big|_{p=p_1+p_2}\nn\\
&\qquad\qquad\quad + 2  \left(D - p \cdot \partial_p \right)
  F(p)\Big|_{p=p_1+p_3} + 2  \left(D - p \cdot \partial_p \right)
  F(p)\Big|_{p=p_2+p_3} \nn\\
&\qquad\qquad + \int_q (D-q \cdot \partial_q) R(q) \cdot h(q)^2\nn\\
&\qquad\qquad\quad \times \left( (N+4) h(q+p_1+p_2) + 2 h(q+p_1+p_3)
  + 2 h (q+p_2+p_3) \right) \Big\rbrace \Bigg]
\end{align}
On the other hand the number operator, defined by
\begin{equation}
\N (0) \equiv - e^{S} \int_q K(q) \frac{\delta}{\delta \phi^I (q)}
\left( \op{\phi^I (q)} e^{-S}\right)\,,
\end{equation}
is calculated as
\begin{align}
\N (0)
&= \int_p \frac{1}{2} \op{\phi^I (p)} \op{\phi^I (-p)} \Big[ 2 p^2 +
  (N+2) \lambda \left( - 2 v_2 + \int_q R(q) h(q)^2 \right)\nn\\
&\quad + \lambda^2 (N+2) \lb - 6 G(p) + 2 (N+2) v_2 \int_q R(q) h(q)^3 - 6
  \int_q R(q) h(q)^2 F(q+p) \rb \Big]\nn\\
&\quad + \frac{1}{2} \int_{p_1,\cdots,p_4} \frac{1}{2} \op{\phi^I
  (p_1)} \op{\phi^I (p_2)} \frac{1}{2} \op{\phi^J (p_3)} \op{\phi^J
  (p_4)}\, \delta \left(\sum_{i=1}^4 p_i\right)\nn\\
&\qquad \times \Big[ 4 \lambda - 4 \lambda^2 \lb
(N+4) F(p_1+p_2) + 2 F(p_1+p_3) + 2 F(p_2+p_3) \rb \nn\\
&\qquad\quad - 2 \lambda^2 \int_q R(q) h(q)^2 \big\lbrace (N+4)
  h(p+p_1+p_2) \nn\\
  &\qquad\qquad\qquad\qquad\qquad + 2 h(p+p_1+p_3) + 2 h(p+p_2+p_3)
    \big\rbrace \Big] 
\end{align}

Using
\begin{equation}
    f(q) = \left(q \cdot \partial_q + 2 \right) h(q) = (2 - q
    \cdot \partial_q) R(q) \cdot h(q)^2
\end{equation}
and the equations satisfied by $F$ and $G$
\begin{subequations}
\begin{align}
\left( p \cdot \partial_p + \ep \right) F(p) 
&= \int_q f(q) \cdot \left(h(q+p)
  - h(q) \right)\\
\left( p \cdot \partial_p - 2 + 2 \ep \right) G (p)
&= \frac{2}{3} v_2 \int_q  f(q) 
  \cdot h(q) + \eta \,p^2 + \int_q  f(q) \cdot F(q+p)
\end{align}
\end{subequations}
we obtain
\begin{align}
&  \Theta (0) + \left(\frac{D-2}{2} + \gamma_N^{(2)} \lambda^2 \right)
  \N (0)\nn\\
&= \left(\ep \lambda + \beta_N^{(1)} \lambda^2 \right) \Bigg[
  \int_p \frac{1}{2} \op{\phi^I (p)} \op{\phi^I (-p)} \, (N+2) v_2\nn\\
  &\quad - \frac{1}{2} \int_{p_1,p_2,p_3, p_4} \frac{1}{2} \op{\phi^I
    (p_1)} \op{\phi^I (p_2)} \frac{1}{2} \op{\phi^J (p_3)} \op{\phi^J
    (p_4)} \,\delta \left(\sum_{i=1}^4 p_i \right)\Bigg]
\end{align}
where we have dropped $\ep \lambda^2 G(p)$ and $\ep \lambda^2 F(p)$,
which are terms of order $\ep^3$.  This vanishes at the fixed point,
where
\[
  \ep \lambda + \beta_N^{(1)} \lambda^2 = 0,
\]
to order $\ep^2$.

\subsection*{Correlation functions}

In the previous section we saw how the fixed-point Wilson action gives
the correlation functions.  Similarly, the coefficient functions
$c_{\mu\nu, 2} (p_1, p_2)$ and $c_{\mu\nu, 4} (p_1,p_2; p_3,p_4)$ give
the 1PI correlation functions of the energy-momentum tensor at $p=0$:
\begin{align}
 \vev{\Theta_{\mu\nu} (0) \phi^I (p) \phi^J (q)}^{1PI}
    &= p^{2-\eta} q^{2 -\eta} \vev{\Theta_{\mu\nu} (0) \phi^I (p)
    \phi^J (q)}\nn\\
  &= \delta (p+q) \delta^{IJ} \lim_{t \to \infty} e^{(-2 + \eta) t}
    c_{\mu\nu,2} (p e^t, - p e^t)
\end{align}
and
\begin{align}
&\vev{\Theta_{\mu\nu} (0) \phi^I (p_1) \phi^J (p_2) \phi^K (p_3) \phi^L
  (p_4)}^{1PI} \nn\\
&\qquad= \prod_{i=1}^4 p_i^{2-\eta} \cdot
                 \vev{\Theta_{\mu\nu} (0) \phi^I (p_1) \phi^J (p_2)
                 \phi^K (p_3) \phi^L (p_4)} \nn\\
&\qquad= \delta \left( \sum_{i=1}^4 p_i \right) \lim_{t \to \infty} e^{(-\ep+4\eta)t}
  \left[ \delta^{IJ}   \delta^{KL} c_{\mu\nu,4}
  (p_1 e^t,p_2 e^t; p_3 e^t,p_4 e^t) \right.\nn\\
&\qquad\qquad \left.+ \delta^{IK} \delta^{JL} c_{\mu\nu,4} (p_1
  e^t,p_3 e^t; p_2 e^t,p_4 e^t)  + \delta^{IL} \delta^{JK} 
 c_{\mu\nu,4} (p_1 e^t,p_4 e^t; p_2 e^t,p_3 e^t) \right]
\end{align}
We obtain the two-point function as
\begin{align}
  \lim_{t\to\infty} e^{(-2 + \eta) t} c_{\mu\nu,2} (p e^t, -p e^t)
  &=  \lim_{t\to\infty} \lb\left(1 + \eta \, t\right) \left( - p^2
    \delta_{\mu\nu} + 2 p_\mu p_\nu \right) \right.\nn\\
  &\left.\quad + \lambda^2 (N+2) 3 e^{- 2
    t} \left(
    \delta_{\mu\nu} G( p e^t) - 2 p_\mu p_\nu e^{2t} G' (p e^t) \right)
    \rb\nn\\
  &= p^{-\eta} \left( - p^2 \delta_{\mu\nu} + 2 p_\mu p_\nu\right)
\end{align}
where we have used the asymptotic form
\begin{equation}
G (p) - 2 p_\mu p_\nu G' (p) \overset{p \to \infty}{\longrightarrow}
\frac{1}{12 (4 \pi)^4} \left( p^2 \delta_{\mu\nu} - 2 p_\mu p_\nu
\right) \ln p^2
\end{equation}
We obtain the four-point function as
\begin{align}
  &  \lim_{t \to \infty} e^{(- \ep + 4 \eta) t} c_{\mu\nu,4} (p_1 e^t, p_2 e^t;
    p_3 e^t, p_4 e^t)\nn\\
  &= \lambda \lim_{t \to \infty} (1 - \ep \, t) \Bigg[
    \delta_{\mu\nu} \big\lbrace - 1 \nn\\
  &\qquad\qquad + \lambda \left( (N+4) F\left((p_1+p_2)e^t\right) + 2
    F\left((p_1+p_3)e^t\right) + F\left((p_2+p_3)e^t\right) \right)
    \big\rbrace \nn\\
  &\,\,- \lambda \lb (N+4) \frac{(p_1+p_2)_\mu
    (p_1+p_2)_\nu}{(p_1+p_2)^2} + \frac{(p_1+p_3)_\mu
    (p_1+p_3)_\nu}{(p_1+p_3)^2} + \frac{(p_2+p_3)_\mu
    (p_2+p_3)_\nu}{(p_2+p_3)^2} \rb \Bigg]\nn\\
    &= - \lambda \dd_{\mu\nu}\left[ 1+ \frac{\eps}{N+8} \ln \lb
      (p_1+p_2)^{N+4}(p_1+p_3)^2(p_2+p_3)^2\rb\right]
\end{align}
where we have kept only the logarithms of momenta at order $\ep^2$.

\section{Summary and Conclusions:}

In this paper we have studied some aspects of the $O(N)$ model using the Exact RG formalism. We have done two things:

1) We have constructed the  Wilson action for the $O(N)$ model at the Wilson Fisher fixed point in $4-\eps$ dimensions up to order $\eps^2$. This is done by solving the fixed point equation, order by order in $\eps$.  Some correlation functions have also been calculated.

2) We have constructed the energy momentum tensor for this theory. This is done by solving the Ward Identity for diffeomorphism invariance.
The traceless-ness of the energy momentum tensor implies that the Wilson action is scale and conformal invariant. It is important to note that all this is in the presence of a {\em finite cutoff} $\lm$.

As mentioned in the introduction, one of the motivations for this construction is the use the ideas in \cite{Sathiapalan:2017frk,Sathiapalan:2019zex} and construct the AdS action corresponding to this CFT. A related problem is to construct the AdS action for sources for composite operators such as $\phi^i \phi^i$. Even more interesting would be to study the massless spin 2 field that would be the source for the energy momentum tensor. This would give dynamical gravity in the bulk as a consequence of Exact RG in the boundary by a direct change of variables similar to what was done for the scalar field in \cite{Sathiapalan:2017frk,Sathiapalan:2019zex}.

\begin{appendices}

\section{Fixed Point Action}

\subsection{Evaluation of $U_4$}\label{evaluatev4}
\begin{figure}[b]
   \centering
  \begin{tikzpicture}
  \begin{feynman}
    	\vertex (e) at (1,0);
      \vertex (m) at ( 0, 0);
      \vertex (n) at ( 3,0);
      \vertex (x) at (-2,0){\(\phi^I\)};
      \vertex (y) at (5,0);
      \vertex (a) at (-1,-2){\(\phi^J\)};
      \vertex (b) at ( 5,0){\(\phi^J\)};
      \vertex (c) at (-1, 2){\(\phi^I\)};
      \diagram* {
        (m) -- [ edge label= $p$, near end](n),
      	(m)-- [edge label =$p_2$](x),
        (a) -- [edge label'=$p_3$](m),
        (c) -- [edge label = $p_1$](m),
        (b) -- [near start,edge label = $p_4$, edge label'=$-p$, near end] (n),
        (n) -- [out=120, in=60, min distance= 3cm] n,
        };
    \end{feynman}
  \end{tikzpicture}
  \caption{Type I diagram }
  \label{type1diagram}
\end{figure}
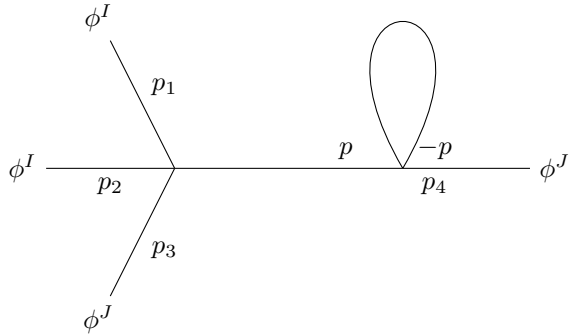

We need to solve
\begin{align} \label{952.1}
\nonumber &\bigg[\bigg(4-D - \sum_{i=1}^4 p_i \frac{d}{d p_i}\bigg) +\sum_{j=1}^4 2 K'(p_j^2) U_2^{(1)}(p_j)\bigg]\frac{1}{8} U_4 (p _1, p _2; p_3,p_4)\\ 
=\nonumber & \int \Dp K'(p^2)) \frac{1}{48}\bigg \lbrace  6N U_6 (p _1, p _2;p_3,p_4; p,-p)+ 12 U_6 (p _1, p ;p_2,-p;p_3,p_4)+12 U_6(p_1,p_2;p_3,p;p_4,-p)\bigg \rbrace\\
= &\nonumber \int \Dp K'(p^2))\bigg \lbrace -\frac{(N+2)}{8} \bigg( h(p_1) +h(p_2)+ h(p_3)+h(p_4)\bigg) \\
&-\frac{(N+4)}{4}\bigg(h(p+p_1+p_2) + 2 h(p+p_1+p_3) +2 h(p+p_1+p_4)\bigg)\bigg \rbrace
\end{align}


where
\begin{equation}\label{type1}
\int \Dp K'(p^2)) \bigg \lbrace -\frac{(N+2)}{8} \bigg( h(p_1) +h(p_2)+ h(p_3)+h(p_4)\bigg ) \bigg \rbrace 
\end{equation}
corresponds to the kind of diagrams shown in \ref{type1diagram}. Here the external loop does not involve momenta $p_i+p_j$. We will call it Type I diagrams. Considering only leading order terms in $p_j^2$ the contribution from type I diagram in \eqref{952.1} is 


\begin{align} \label{type1value}
= -\frac{N+2}{8} \frac{\lambda^2}{16\pi^2} ~ 4 K^\prime(p_j^2)\bigg\vert_{p_j=0}
\end{align}
Now consider the second term in L.H.S of \eqref{952.1}. In the limit of small external momenta after putting the value of $U_2^{(1)}(p)=-\frac{N+2}{2}\frac{\lambda}{16\pi^2}$( as we are considering terms of $\mathcal{O}(\epsilon^2)$ we have put D=4 to find $U_2^{(1)}$) we get
\begin{align}
\nonumber -&\sum_{j=1}^4 2 K^\prime(p_j^2)\bigg \vert_{p_j\rightarrow 0}\frac{\lambda}{16\pi^2}\frac{N+2}{2}\frac{1}{8} V_4 (p _1, p _2; p_3,p_4)\\ 
=& -4 K^\prime(p_j^2)\bigg\vert_{p_j\rightarrow 0}\frac{\lambda^2}{16\pi^2}\frac{N+2}{8}
\end{align}
This  cancels exactly with \eqref{type1value}.

\begin{figure}
   \centering
  \begin{tikzpicture}
  \begin{feynman}
  \vertex (r) at (0,0);
      \vertex (m) at (-2, 0);
      \vertex (n) at ( 2,0);
      \vertex (x) at (-4,0);
      \vertex (y) at (4,0);
      \vertex (a) at (-3,-2);
      \vertex (c) at (-3, 2); 
      \vertex (d) at (3,2);
      \vertex (e) at (3,-2); 
      \node at (-3.5,2) {$\phi^I$};
      \node at (3.5,-2) {$\phi^J$};
      \node at (3.5,-2) {$\phi^J$};
      \node at (3.5,2) {$\phi^I$};
      \node at (-3.5,-2){$\phi^K$};
      \node at (4.5,0){$\phi^J$};
      \node at (-4.5,0){$\phi^K$};
      \diagram* {
        (c) -- [edge label= $p$](m),
        (d) -- [edge label' = $-p$](n),
        (x)--  [edge label =$p_1$](m)--(r)-- (n) -- [edge label=$p_3$](y),
        (m) -- [edge label =$p_2$](a),
        (n) -- [edge label'= $p_4$](e),
        };
       \end{feynman}
       \draw [dashed] (c) .. controls (0,3).. (d);
    
  \end{tikzpicture}
  \caption{Type II diagram }
  \label{type2diagram}
\end{figure}
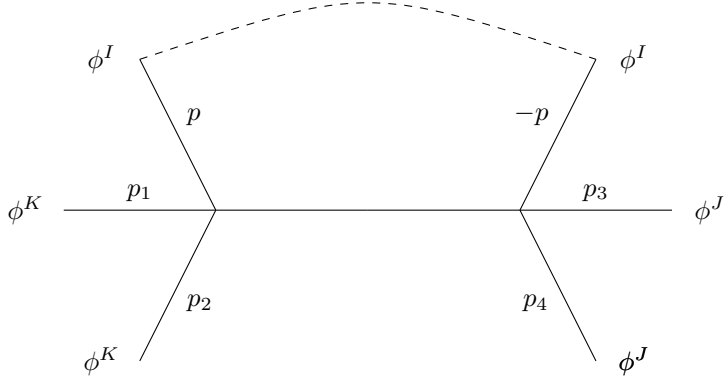

Similarly in \eqref{952.1} the term
\begin{align}
\int \Dp K'(p^2)\bigg \lbrace-\frac{(N+4)}{4}\bigg(h(p+p_1+p_2) + 2 h(p+p_1+p_3) +2 h(p+p_1+p_4)\bigg)\bigg \rbrace
\end{align}

corresponds to the kind of diagram shown in \ref{type2diagram}. We will call it Type II diagram. In the limit $p_i \rightarrow 0$ the above term becomes

\begin{align*}
& \lambda^2~\frac{(N+8)}{4}\frac{1}{16\pi^2}\int_0^\infty dp^2 K^\prime(p^2) \Big(K(p^2)-K(0)\Big)\\
=&\lambda^2 \frac{(N+8)}{4}\frac{1}{16\pi^2}\int_0^\infty dp^2 \bigg \lbrace \frac{1}{2}\frac{d(K^2)}{dp^2}-K(0)K^\prime(p^2) \bigg \rbrace 
\end{align*}
Using $K(\infty)=0$ and $K(0)=1$, this integral gives $\frac{1}{2}$. Equating this contribution with $\epsilon \frac{\lambda}{4!}$ from L.H.S of \eqref{952.1} we obtain
\[
\frac{1}{8} (4-D) \lambda = \frac{N+8}{8}\frac{\lambda^2}{(4\pi)^2}
\]
Thus in addition to the trivial fixed point $\lambda=0$, we have a non trivial fixed point:
\be
 \boldmath \lambda = (4-D) \frac{16\pi^2}{N+8}
\ee

\subsection{Solving for $\tilde U_4$}\label{u4}
\label{U4cal}

$\tilde{U}_4$ will have contribution from both type I and II diagram explained above. We write 
\begin{align*}
\tilde{U}_4=\tilde{U}_4^{I}+\tilde{U}_4^{II}
\end{align*}
according to contributions from type I(II) diagrams.
\newline
 (We shall set $D=4$ while evaluating  integrations in those terms that are already of $\mathcal{O}(\epsilon^2)$.)



\paragraph{Type I diagram}
In \eqref{952.1} the first term on the LHS and the first terms on the RHS (Type I) cancel only in leading order. In general their difference  is
\[
\lambda^2\frac{N+2}{8}\times \frac{1}{(4\pi)^2}\int_0^\infty dp^2 K^\prime(p^2)\Bigg[ \sum _j \frac{K(p_j^2)-K(0)}{p_j^2} - K'(p_j^2)\Bigg]
\]
Taylor expanding we find
\[
\lambda^2\frac{N+2}{8}\times \frac{1}{(4\pi)^2}\int dp^2 K^\prime(p^2)K''(0)\hf \sum _j p_j^2 \equiv c \sum_j p_j^2
\]
This is a contribution to $\tilde U_4(p_1,p_2;p_3,p_4)$ that we can call $\Delta U_4^I(p_1,p_2;p_3,p_4)$.
Consider a type I graph where the line at one end has $p_1$ and lines with momenta $p_2,p_3,p_4$ are at the other end. This corresponds to the term
\[
\lambda^2\frac{N+2}{8}\times \frac{1}{(4\pi)^2}\int dp^2 K^\prime(p^2)K''(0)\hf p_1^2 \equiv c  p_1^2
\] 
when contracted in a loop in order to contribute to $\tilde U_2$, so that say $p_3=-p_4$,  we have $p_2=-p_1$.  It contributes to $\tilde U_2(p_1^2)$ an amount
\begin{align*}
\int dp^2 K'(p^2)\hf \Delta U_4^I(p_1,-p_1,p,-p)= \int dp^2 K'(p^2)\hf c(p_1^2)
=\Big[c \int dp^2 K'(p^2)\Big] p_1^2 \equiv  A p_1^2
\end{align*}
This is just a simple wave function renormalization that does not depend on $p_1$. There is no contribution to the mass.
The same argument applies to all the other permutations of the type I terms.
A simple wave function renormalization $\phi '^2=(1+A) \phi^2$ can ensure the normalization of the kinetic term..
They do not affect the physics or contribute to $\eta$. However, type-I term contributes to sub-leading order term of $m^2$ or $U_2$. 


$\tilde{U}_4^I$ satisfies the following equation:

\begin{align}
-\sum_{i=1}^4p_i\frac{d}{d p_i} \frac{1}{8}\tilde{U}_4^I(p_1,p_2;p_3,p_4)=\lambda^2\frac{N+2}{8}\times \frac{1}{(4\pi)^2}\int_0^\infty dp^2 K^\prime(p^2)\Bigg[ \sum _j \frac{K(p_j^2)-K(0)}{p_j^2} - K'(p_j^2)\Bigg]
\end{align}

The solution is
\begin{subequations}
\begin{align}\label{U4I}
\tilde{U}_4^I(p_1,p_2;p_3,p_4)=&-\lambda^2 \frac{(N+2)}{2} \frac{1}{16\pi^2}\sum_{j=1}^4 \frac{K(p_j^2)-K(0)}{p_j^2}\\
=& \lambda^2 \frac{(N+2)}{2} \frac{1}{16\pi^2}\sum_{j=1}^4 h(p_j)
\end{align}
\end{subequations}

where $K(p)=e^{-p^2}$ is assumed.

\paragraph{Type II Diagram}

%
%
%
%
%
%
%
%
%
%
%

In \eqref{952.1} if we keep terms upto $\mathcal{O}(\epsilon^2)$,
\begin{align} \label{U4tilde}
\nonumber & \frac{1}{8}\bigg[ \sum_{j=1}^4p_j\frac{d}{dp_j}\bigg]\tilde U_4^{II}(p_1,p_2;p_3,p_4)\\
=&\frac{\lambda^2}{4}\int \Dp K'(p^2)\bigg \lbrace (N+4)h(p+p_1+p_2)+2 h(p+p_1+p_3)+2 h(p+p_1+p_4)-(N+8) h(p)\bigg \rbrace
\end{align}
where $h(p)=\frac{K(0)-K(p)}{p^2}$. 
It is to be noted in the momentum independent part $-\epsilon \frac{\lambda}{4!}$ we have written $\epsilon$ in terms of $\lambda$ using the fixed point value of $\lambda$.
\newline
The solution at $\mathcal{O}(\epsilon^2)$, analytic at zero external momenta, is given by
\begin{subequations}
\begin{align}\label{U4II}
& \nonumber \tilde U_4^{II}(p_1,p_2;p_3,p_4)\\
=&-\frac{\lambda^2}{2} \int \Dp h(p)\Big[(N+4)h(p_1+p_2+p)+2 h(p+p_1+p_3)+2 h(p+p_1+p_4)-(N+8)h(p)\Big]\\
=&-\lambda^2 \Big[(N+4)F(p_1+p_2)+2 F(p_1+p_3)+2 F(p_1+p_4)\Big]
\end{align}
\end{subequations}
where $F(q)=\frac{1}{2}\int \Dp h(p)\Big(h(p+q)-h(p)\Big)$.

\subsection{Equation for $\tilde U_2$}
\label{U2cal}

 From \eqref{U2.1} we get
\begin{align}\label{U2.2}
\nonumber 0=&\int  \Dp \Big(-K'(p^2)\Big) \times\\
 \nonumber \bigg \lbrace &\frac{1}{8}\Big[4N\tilde{U}_4^I(p_1,-p_1;p,-p)+4N \tilde{U}_4^{II}(p_1,-p_1;p,-p)+ 8\tilde{U}_4^{I}(p_1,p;-p_1,-p)+8\tilde{U}_4^{II}(p_1,p;-p_1,-p)\Big]\\
- &v_2^{(1)}(p)v_2^{(1)}(p)\dd^D(p - p_1)\bigg \rbrace- \frac{\eta}{2}p_1^2+ \tilde{U}_2(p_1)-p_1^2\frac{d\tilde{U}_2(p_1)} {dp_1^2}
\end{align}
From\eqref{U4I}
\begin{align}
\nonumber &\frac{1}{8}\bigg \lbrace 4N\tilde{U}_4^I(p_1,-p_1;p,-p)+8\tilde{U}_4^I(p_1,p;-p,-p_1)\bigg \rbrace\\
= & \frac{1}{2}(N+2)^2\frac{\lambda^2}{16\pi^2}\bigg \lbrace h(p)+ h(p_1) \bigg \rbrace
\end{align}
and  from \eqref{U4II}
\begin{align}
\nonumber &\frac{1}{8}\bigg \lbrace 4N\tilde{U}_4^{II}(p_1,-p_1;p,-p)+8\tilde{U}_4^{II}(p_1,p;-p,-p_1)\bigg \rbrace\\
= & -\frac{3\lambda^2}{2}(N+2) \int_{ r} \bigg \lbrace h( r)\Big[h( r+ p_1+ p)-h( r)\Big] \bigg \rbrace
\end{align}
If we decompose $\tilde{U}_2$ in two parts namely $\tilde{U}_2^I$ and $\tilde{U}_2^{II}$ respectively, in the following way,

1.
\begin{align}\label{tildeU2.1}
\tilde{U}_2^I(p_1)-p_1^2\frac{d\tilde{U}_2^I(p_1)} {dp_1^2}= \int \Dp K^\prime(p^2) \frac{1}{2}(N+2)^2\frac{\lambda^2}{16\pi^2} h(p_1) - \big(U_2^{(1)}\big)^2 K^\prime(p_1^2)
\end{align}
which gives
\begin{align}
\tilde{U}_2^I(p_1)= -\frac{\lambda^2}{(16\pi^2)^2}\frac{(N+2)^2}{4} h(p_1)
\end{align}
2.
\begin{align}\label{tildeU2.2}
& \nonumber -2\tilde{U}_2^{II}(p_1)+2p_1^2\frac{d\tilde{U}_2^{II}(p_1)} {dp_1^2}\\
=&  -6\lambda^2(N+2)\int \Dp \Big(- K'( p^2)\Big)F(p_1+p)+ (N+2)^2\frac{\lambda^2}{16\pi^2}\int \Dp\Big(-K^\prime(p^2)\Big)h(p)-\eta ~p_1^2
\end{align}
which gives
\begin{align}
\tilde{U}_2^{II}(p_1)=~& p_1^2\int_{p^2 = 0}^{p_1^2} dp^2 \frac{\int \frac{d^D q}{(2\pi)^D}\Big\lbrace-6\lambda^2(N+2)(- K'( q^2))F(p+q)\Big\rbrace-\eta~p^2}{2p^4}- \frac{(N+2)^2}{4}\frac{\lambda^2}{(16\pi^2)^2}
\end{align}
The second term in the expression of $\tilde{U}_4^{II}$ is evaluated using $K(p)=e^{-p^2}$.

\vspace{0.05 in}
Hence The full expression of $\tilde{U}_2(p_1)$ is given by
\begin{boldmath}
\begin{align}\label{tildeU2full}
\nonumber \tilde{U}_2(p_1)=-&\frac{\lambda^2}{(16\pi^2)^2}\frac{(N+2)^2}{4} h(p_1)\\
+& p_1^2\int_{p^2 = 0}^{p_1^2} dp^2 \frac{\int \frac{d^D q}{(2\pi)^D}\Big\lbrace-6\lambda^2(N+2)(- K'( q^2))F(p+q)\Big\rbrace-\eta~p^2}{2p^4}- \frac{(N+2)^2}{4}\frac{\lambda^2}{(16\pi^2)^2}
\end{align}
\end{boldmath}

\subsection { Expression for $\eta$}\label{ETA}
Only Type II diagrams contribute to $\eta$.
Because we need the external momentum to flow through the loop - to get a momentum dependence in $U_2$. This can happen
only in Type II terms and that too for \textbf{certain contractions}.

\vspace{0.05 in}
(Calculation of this section requires us to go back to bar denoted variable as dimensionless variable. So $p$'s from last section are replaced with $\bp$. )

\vspace{0.05 in}
From \eqref{eta} we have
\be \label{eta1.1}
\frac{\eta}{2} = -\frac{1}{8}\frac{d}{d\bar r^2} \int _{\bar q} K'(\bar q^2)\bigg \lbrace 4N\tilde{U}_4^{II}(\bar q,-\bar q;\bar r,-\bar r)+ 8\tilde{U}_4^{II}(\bar q,\bar r;-\bar r,-\bar q) \bigg \rbrace ~\Bigg|_{\bar r^2=0}
\ee
We can convert differentiation w.r.t $p_j$ into that w.r.t $\Lambda$ , i.e.
\begin{align*}
-\sum_{j=1}^4\bp_j\frac{d}{d\bp_j}=\lm\frac{d}{d\lm}
\end{align*}
 So \eqref{U4tilde} gives following expression for $\tilde{U}_4^{II}$:

\begin{align} \label{U4.1}
 \nonumber \frac{1}{8}&\tilde {U}_4^{II}(\frac{p_1}{\lm},\frac{p_2}{\lm};\frac{p_3}{\lm},\frac{p_4}{\lm})\\
 = &\frac{\lambda^2}{4}\int _0^{\ln \lm}d \ln \lm'~\int_{\bp} K'(\bp^2)
  \bigg[(N+4)h(\bp +\frac{p_1}{\lm'}+\frac{p_2}{\lm'})+ 2 h(\bp +\frac{p_1}{\lm'}+\frac{p_3}{\lm'})+ 2 h(\bp+\frac{p_1}{\lm'}+\frac{p_4}{\lm'})-(N+8)h(\bp)\bigg]
\end{align}
Hence
\begin{align}\label{eta1.2}
\nonumber &\frac{1}{8}\bigg \lbrace  4N\tilde{U}_4^{II}(\bar q,-\bar q;\bar r,-\bar r)+ 8\tilde{U}_4^{II}(\bar q,\bar r;-\bar r,-\bar q)\bigg \rbrace \\
=& \frac{\lambda^2}{4}\int _0^{\ln \lm}d \ln \lm'~\int_{\bp,\bar r} K'(\bp^2)\bigg \lbrace (12N+48)h(\bp +\frac{q}{\lm'}+\frac{r}{\lm'})+(12N+48)h(\bp +\frac{q}{\lm'}-\frac{r}{\lm'})-24(N+2) h(\bp)\bigg \rbrace
\end{align}
So we need to find the coefficient of $\bar r^2$ in $\Big[  h(\bp +\frac{q}{\lm'}+\frac{r'}{\lm'})+h(\bp+\frac{q}{\lm'}-\frac{r'}{\lm'})\Big] $ which is calculated as
\begin{align}\label{eta1.3}
& \nonumber\hf \frac {r^\mu r^\nu}{\lm'^2}\frac{d^2}{dr'^\mu dr'^\nu}\Big[  h(\bp +\frac{q}{\lm'}+\frac{r'}{\lm'})+h(\bp+\frac{q}{\lm'}-\frac{r'}{\lm'})\Big]\Bigg|_{r'=0}\\
=\nonumber &\frac {r^\mu r^\nu}{\lm'^2}\bigg(\frac{d^2}{d\bar r'^\mu d\bar r'^\nu}  h(\bp +\frac{q}{\lm'}+\bar r'\bigg)\Bigg|_{\bar r'=0}\\
=& \nonumber \frac{\bar r^2}{4}\bigg( \frac{d^2}{d\bar r^\mu d\bar r_\mu} h(\bp + \frac{ q}{\Lambda^\prime} +\bar r\bigg)\Bigg|_{r=0}\\
=& \nonumber -\frac{\bar r^2}{4}\frac{d^2}{d \bar r^\mu d \bar r_\mu}\frac {K(\bar r^2)-1}{\bar r^2}\Bigg|_{\bar r= \bp+ \frac{q}{\Lambda^\prime}}\\
=&  \bar r^2 K''((\bp+\frac{q}{\lm'})^2)
\end{align}
where we have used  the facts: in 4 dimensions $(\frac{d}{dp_\mu}\frac{1}{p^2})=\delta^4(p)$ and $K(0)=1$.


From \eqref{eta1.1},\eqref{eta1.2} and \eqref{eta1.3}  we get
\be
\frac{\eta}{2}=3\lambda^2 (N+2)\int _{\bar q} K'(\bar q^2)\int _0^{\ln \lm}d \ln \lm'~(\frac{\lm}{\lm'})^2\int_{\bp} K'(\bp^2)K''((\bp + \frac{ q}{\lm'})^2)
\ee

{\bf Evaluation of integral:}
Let us use $\bar q'= \frac{q}{\lm'}$ and $\lm'$ as variables of integration, rather than $\bar q=\frac{q}{\lm}$ and $\lm'$. So change variables: 
\[
\bar q= \bar q' \frac{\lm'}{\lm} ~;~~~\bar q^2 = \bar q'^2\Big(\frac{\lm'}{\lm}\Big)^2 ~;~~~\int d^4\bar q = \int d^4\bar q \Big(\frac{\lm'}{\lm}\Big)^4
\]
to get
\[
\frac{\eta}{2}=-3\lambda^2(N+2)\int _0^{\ln \lm}d \ln \lm'~\int _{\bar q'} \Big(\frac{\lm'}{\lm}\Big)^{-2}K'(\bar q'^2)\Big(\frac{\lm'}{\lm}\Big)^2 \int_{\bp} K'(\bp^2)K''((\bp + \frac{ q}{\lm'})^2)
\]
Using $ K'(\bar q'^2)= \frac{dK}{d\lm'}\frac{d\lm'}{d\bar q'^2}= -\frac{\lm'}{2\bar q'^2}\frac{dK}{d\lm'}$ we get
\[
\frac{\eta}{2}=-3\lambda^2(N+2)\int _0^{ \lm}d \lm'~\frac{dK}{d\lm'}\int _{\bar q'}\frac{1}{2\bar q'^2}\int_{\bp} K'(\bp^2)K''((\bp + \bar q')^2)
\]
Since $\bar q'$ is an independent variable we can write this as
\[
\frac{\eta}{2}=-3\lambda^2(N+2)\int _{\bar q'}\int _0^{ \lm}d \lm'~\frac{dK}{d\lm'}\frac{1}{2\bar q'^2}\int_{\bp} K'(\bp^2)K''((\bp + \bar q')^2)
\]
The integral over $\bp$ is a function of $\bar q'$ and not $\lm'$. So we can do the $\lm'$ integral easily. Using $K(\infty)=0$ we get
\[
\frac{\eta}{2}=-\frac{3\lambda^2}{2}(N+2)\underbrace{\int _{\bar q'}~K(\bar q'^2)\frac{1}{\bar q'^2}\int_{\bp} K'(\bp^2)K''((\bp + \bar q')^2)}_{-\frac{\pi^4}{6(2\pi)^8}}= \frac{1}{4}\lambda^2(N+2)\frac{1}{(16\pi^2)^2}
\]
The integral underbraced above is calculated to give $-\frac{\pi^4}{6(2\pi)^8}$ for $K(x)=e^{-x}$. But it can be shown to give identical result for any smooth $K(x)$ \cite{Liu-integral}.
Using $\lambda = \frac{16\pi^2}{N+8}\epsilon$ we can write the anomalous dimension as:
\be \boldmath
\frac{\eta}{2} =\frac{1}{4}\lambda^2(N+2) \frac{1}{(16\pi^2)^2}=\frac{N+2}{(N+8)^2}\frac{\epsilon^2}{4}
\ee

\section{ Asymptotic behaviors of $F (p)$ and $G(p)$}
\label{asymptotic}

The function $F(p)$ is defined by
\begin{equation}
\left( p \cdot \partial_p + \ep \right) F(p) = \int_q f(q) \Big(
  h(q+p) - h(q) \Big)
\end{equation}
For large $p$, we obtain an equation satisfied by the
asymptotic form $F_{\mathrm{asymp}} (p)$:
\begin{equation}
\left( p \cdot \partial_p + \ep \right) F_{\mathrm{asymp}} (p) = -
\int_q f(q) h(q) = - \frac{1}{(4 \pi)^2} + \mathrm{O} (\ep)
\end{equation}
This implies
\begin{equation}
F_{\mathrm{asymp}} (p) = - \frac{1}{\ep} \int_q f(q) h(q) + C_F (\ep)
p^{-\ep}
\end{equation}
where $C_F(\ep)$ is independent of $p$.  Since $F (p)$ is finite in
the limit $\ep \to 0+$, we must find
\begin{equation}
C_F (\ep) = \frac{1}{\ep} \frac{1}{(4\pi)^2} + \cdots
\end{equation}
Hence, expanding in $\ep$, we obtain
\begin{equation}
F_{\mathrm{asymp}} (p) = - \frac{1}{(4 \pi)^2} \ln p + \mathrm{const}
+ \mathrm{O} (\ep)
\end{equation}

We next consider $G(p)$ satisfying
\begin{equation}
\left( p \cdot \partial_p - 2 + 2 \ep \right) G(p) = \int_q f(q)
F(q+p) +  2 v_2 \int_q f(q) h(q) + \eta^{(2)} p^2
\end{equation}
where
\begin{equation}
\eta^{(2)} = - \frac{d}{dp^2} \int_q f(q) F(q+p)\Big|_{p=0} =
\frac{1}{6 (4\pi)^4} + \mathrm{O} (\ep)
\end{equation}
The asymptotic form $G_{\mathrm{asymp}} (p)$ satisfies
\begin{equation}
\left( p \cdot \partial_p - 2 + 2 \ep\right) G_{\mathrm{asymp}} (p) =
\eta^{(2)} p^2
\end{equation}
This gives
\begin{equation}
G_{\mathrm{asymp}} (p) = \frac{1}{2 \ep} \eta^{(2)} p^2 + C_G (\ep)
p^{2 - 2 \ep}
\end{equation}
Since $G (p)$ is finite as $\ep \to 0+$, we obtain
\begin{equation}
C_G (\ep) = - \frac{1}{\ep} \frac{1}{12 (4\pi)^4} + \cdots
\end{equation}
Hence,
\begin{equation}
G_{\mathrm{asymp}} (p) = p^2 \left( \frac{1}{6 (4 \pi)^4}  \ln p +
\mathrm{const} \right) + \mathrm{O} (\ep)
\end{equation}

\end{appendices}

\end{document}